\documentclass[12pt,preprint]{aastex}

\usepackage{rotating}
\usepackage{amssymb}
\usepackage{longtable}
\usepackage{booktabs}

\shorttitle{Statistical Study of SADs}
\shortauthors{Savage and McKenzie}

\begin{document}


\title{QUANTITATIVE EXAMINATION OF A LARGE SAMPLE OF SUPRA-ARCADE DOWNFLOWS IN ERUPTIVE SOLAR FLARES}


\author{$^{1,2}$Sabrina L. Savage \& $^{2}$David E. McKenzie}
\affil{$^{1}$NASA/Goddard Space Flight Center (Oak Ridge Associated Universities), 8800 Greenbelt Rd Code 671, Greenbelt, MD  20771, USA} 
\affil{$^{2}$Department of Physics, Montana State University, PO Box 173840, Bozeman, MT  59717-3840, USA}


\begin{abstract}

Sunward-flowing voids above post-coronal mass ejection (CME) flare arcades were first discovered using the soft X-ray telescope (SXT) aboard \textit{Yohkoh} and have since been observed with \textit{TRACE} (extreme ultra-violet (EUV)), \textit{SOHO}/LASCO (white light), \textit{SOHO}/SUMER (EUV spectra), and \textit{Hinode}/XRT (soft X-rays (SXR)).  Supra-arcade downflow (SAD) observations suggest that they are the cross-sections of thin flux tubes retracting from a reconnection site high in the corona.  Supra-arcade downflowing loops (SADLs) have also been observed under similar circumstances and are theorized to be SADs viewed from a perpendicular angle.  Previous studies have presented detailed SAD observations for a small number of flares.  In this paper we present a substantial SADs and SADLs flare catalog.  We have applied semi-automatic detection software to several of these events to detect and track individual downflows thereby providing statistically significant samples of parameters such as velocity, acceleration, area, magnetic flux, shrinkage energy, and reconnection rate.  We discuss these measurements, how they were obtained, and potential impact on reconnection models. 

\end{abstract}


\section{\label{sadsiisec:intro}Introduction}

\begin{figure}[!ht] 
\begin{center}

\includegraphics[width=.55\textwidth]{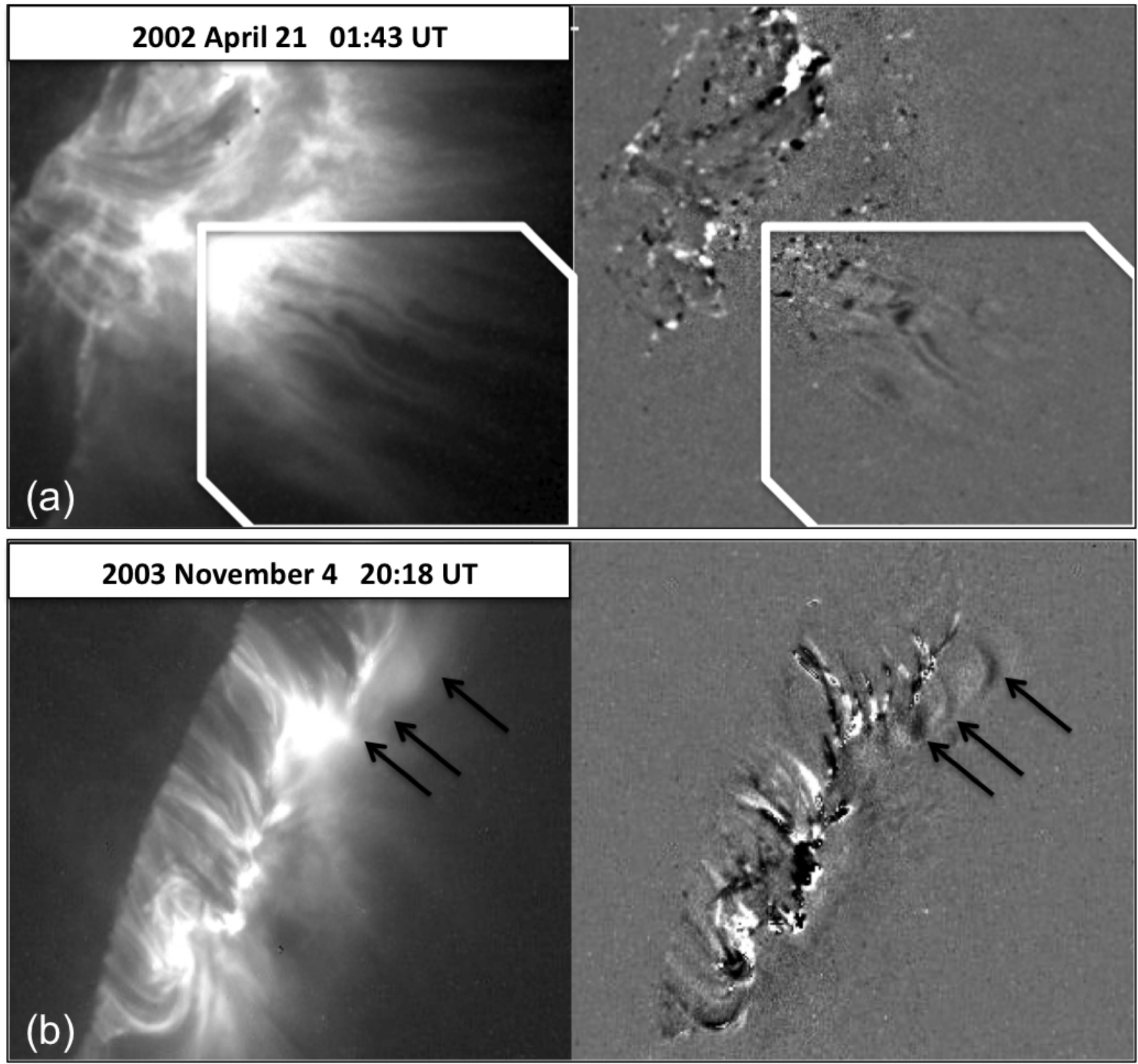}

\caption{(a) Example image from the 2002 April 21 TRACE flare showing supra-arcade downflows (SADs) enclosed within the white box.  (b)  Example image from the 2003 November 4 flare with supra-arcade downflowing loops (SADLs) indicated by the arrows.  The left panel of each set is the original image.  The right panel has been enhanced for motion via run-differencing and scaled for contrast.}
\label{sads_sadls_example}
\end{center}
\end{figure}

Long duration flaring events are often associated with downflowing voids and/or loops in the supra-arcade region (see Figure~\ref{sads_sadls_example} for example images) whose theoretical origin as newly reconnected flux tubes has been supported by observations (\citeauthor{mckenzie-hudson_1999}~\citeyear{mckenzie-hudson_1999}; \citeauthor{mckenzie_2000}~\citeyear{mckenzie_2000}; \citeauthor{innes-mckenzie-wang_2003a}~\citeyear{innes-mckenzie-wang_2003a}; \citeauthor{asai_2004}~\citeyear{asai_2004}; \citeauthor{sheeley-warren-wang_2004}~\citeyear{sheeley-warren-wang_2004}; \citeauthor{khan-bain-fletcher_2007}~\citeyear{khan-bain-fletcher_2007}; \citeauthor{reeves-seaton-forbes_2008}~\citeyear{reeves-seaton-forbes_2008}; \citeauthor{mckenzie-savage_2009}~\citeyear{mckenzie-savage_2009}; \citeauthor{savage_2010}~\citeyear{savage_2010}).

\begin{figure}[!ht] 
\begin{center}

\includegraphics[width=0.9\textwidth]{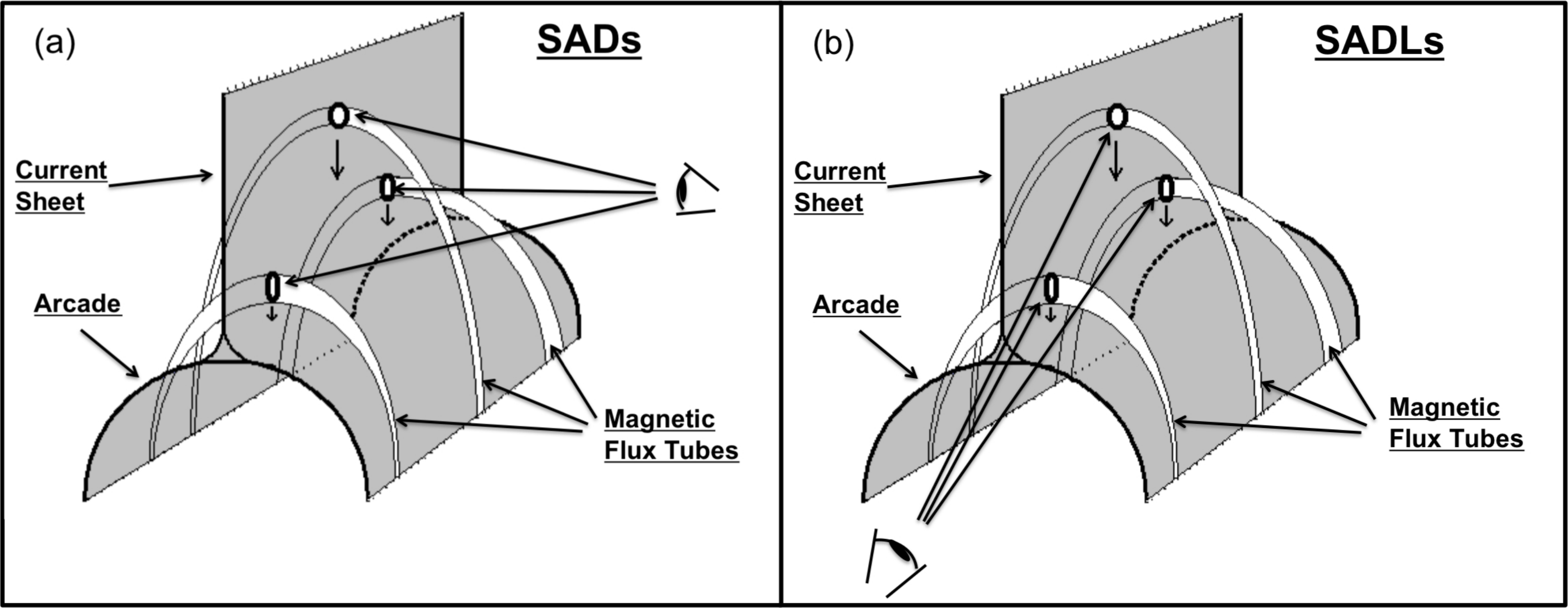}

\caption{(a) Cartoon depiction of supra-arcade downflows (SADs) resulting from 3-D patchy reconnection. Discrete flux tubes are created, which then individually shrink, dipolarizing to form the post-eruption arcade. (b) Cartoon depiction of supra-arcade downflowing loops (SADLs) also resulting from 3-D patchy reconnection.  Note that the viewing angle, indicated by the eye position, is perpendicular to that of SADs observations.}

\label{sads_sadls_diagram_eyes_ch4}
\end{center}
\end{figure}

The downflowing voids, (a.k.a. supra-arcade downflows (SADs) -- Figure~\ref{sads_sadls_example} (a)), differ in appearance from downflowing loops (a.k.a. supra-arcade downflowing loops (SADLs) -- Figure~\ref{sads_sadls_example} (b)); however, the explanation for this can be derived simply from observational perspective.  If the loops are viewed nearly edge-on as they retract through a bright current sheet, then SADs may represent the cross-sections of the SADLs (see Figure~\ref{sads_sadls_diagram_eyes_ch4}).  Since neither SADs nor SADLs can be observed 3-dimensionally by an independent imaging instrument, proving this hypothetical connection is not possible with a single image sequence.  However, their general bulk properties, such as velocity, size, and magnetic flux, can be measured and should be comparable if this scenario is correct.  Moreover, measuring these parameters for a large sample of SADs and SADLs yields constraints that are useful for development of numerical models/simulations of 3D magnetic reconnection in the coronae of active stars.

\begin{figure}[!ht] 
\begin{center}

\includegraphics[width=0.6\textwidth]{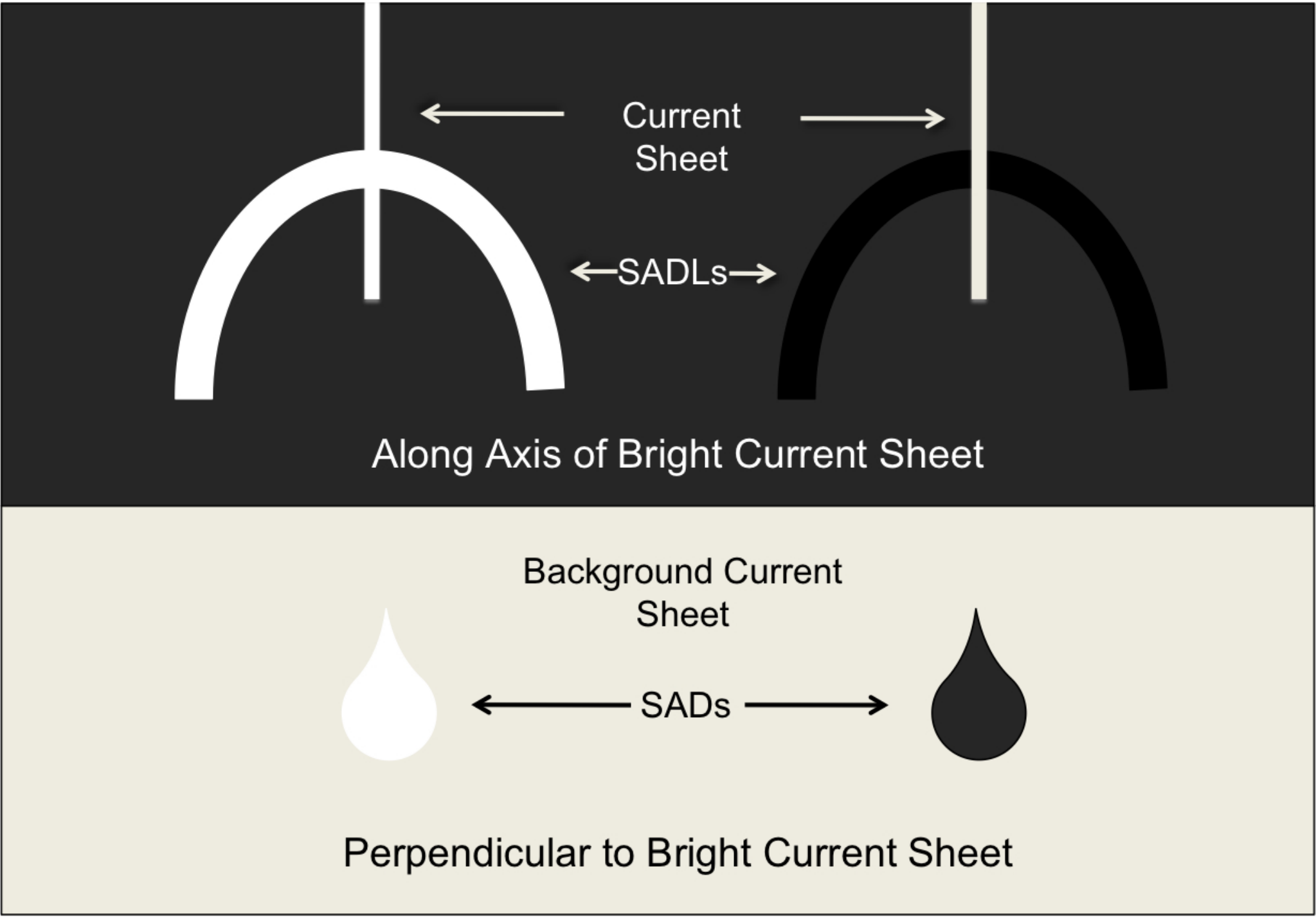}

\caption{Diagram depiction illustrating the possible reason for the lack of bright SADs despite observations of bright SADLs with a snapshot of a loop shrinking through a bright current sheet with a viewing angle \textit{Top:} along the arcade axis and \textit{Bottom:} perpendicular to the axis.  (Different scales)  The loop is bright on the left of both panels and dark on the right.  Because the current sheet is thin and in a region of low signal, the bright loops are easier to observe along the arcade axis which is opposite for the dark loops.}

\label{cartoon_bright_dark}
\end{center}
\end{figure}

While bright shrinking loops (SADLs) are often observed as well as dark ones when the background has been sufficiently illuminated, bright SADs are rarer.  A possibility for the reason behind this lack of bright SADs observations is given in Figure~\ref{cartoon_bright_dark}.  In order to view SADs, the loops are viewed edge-on to their apex cross-sections as they travel through the bright current sheet which provides a bright background.  Consequently, if the loops are nearly as bright as the current sheet, then they blend in with the background and are difficult to observe.  Conversely, because the current sheet is thin, viewing it edge-on gives the appearance of a bright, thin line surrounded by a dark background against which the bright loops can be seen.  The coronal background itself would need to be sufficiently illuminated in order to observe dark loops as they retract.  This is often not the case, however, because during a flare, the footpoints are so bright that the exposure durations applied to the images are not long enough to result in any significant coronal background signal.  A counter-example is the 2008 April 09 flare described in \cite{savage_2010}.

In this paper we provide a list of 62 flares observed by several instruments containing downflow signatures.  We analyze flows from 35 of these flares and present comparative results of general bulk properties, including magnetic flux and shrinkage energy estimates, from SADs and SADLs in Section~\ref{sadsiisec:analysis}.  These comparisons provide compelling evidence linking SADs to SADLs and constraints on flare magnetic reconnection models.  Possible trends in the data are speculated in Section~\ref{sadsiisec:trends}.  The effect of simple drag on the loops is investigated in Section~\ref{sadsiisec:drag} as a possible reason for the slow downflow speeds.  Some loop cooling observations are presented in Section~\ref{sadsiisec:loop_cooling} relating the appearance of shrinking loops and the brightening of the arcade.   Finally, in Section~\ref{sadsiisec:discussion} we summarize our findings of quantities that are typical of the observed SADs and SADLs, and suggest goals for future models of 3D bursty reconnection.

\subsection{\label{sadsiisec:observations}Observations}

Because SADs and SADLs are located in regions of extremely low coronal emission near bright, dynamic sources, measurements for any one flow are naturally associated with a high degree of error.  Therefore, in order to identify any trends in the data, analysis of several flows from each of many flares must necessarily be performed.  Table~\ref{sads_list_full} contains a list of flares which have been noted to display downflow signatures (i.e. observable SADs,  SADLs, or swaying fan above the arcade as suggested by \cite{khan-bain-fletcher_2007}).  Several of these flares were selected from \cite{khan-bain-fletcher_2007} (see Table~1 therein) and were supplemented from \cite{mckenzie_2000} as well as by personal flare data investigation.  Some flows from the \cite{khan-bain-fletcher_2007} list were excluded from this study if the presence of flows was not confirmed by visual inspection.  

The majority of the flares are from SXT observations.  Under the ``Filter" heading in Table~\ref{sads_list_full}, the ``Q-", ``H-" and the ``F-" preceding the filter indicate whether the images examined are quarter-, half-, or full-resolution, respectively (where Q = 9.8, H = 4.9, \& F = 2.5~arcsec/pix).  The SXT resolution is about 2$-$10 times poorer than either TRACE (0.5~arcsec/pix) or XRT (1~arcsec/pix) (depending on SXT resolution); however, it was operational during two solar maxima (unlike XRT to date which has been operational primarily during an unusually quiet solar minimum) and observed hotter plasma than TRACE.  Having the capacity to observe the hot plasma in the current sheet increases the height of flow observations.  A few of the flares lack GOES assignments either because the soft X-ray (SXR) output was too low or the footpoints were too far beyond the limb to measure any significant signal. 

Note that the TRACE flares in Table~\ref{sads_list_full} are extremely energetic (as indicated by their GOES X1.5, X4.8, \& X28 classifications).  In addition, the TRACE flares are observed with the 195~\AA\ filter which has temperature response peaks both in the .5 $-$ 2~MK and 11 $-$ 26~MK bandpasses.  The high energies result in very high temperature plasma detectable above the underlying post-eruptive arcade.  The increased intensity in the supra-arcade region, presumably within the current sheet \citep{reeves_2010}, provides a bright background against which to observe the dark downflows.

Not all of the flares listed are suitable for tracking flows due to various factors (e.g. cadence, flow visibility, flare position, image quality, etc.).  The last column of Table~\ref{sads_list_full} indicates whether analysis of a flare was performed using the semi-automated routines described in \cite{mckenzie-savage_2009}~(Section~2 therein) or a supplementary manual-tracking routine.  Flows for 35 out of the 62 flares were evaluated.  Table~\ref{sads_list_analyzed} includes a list of the analyzed flares from this study.  Whether the flows were determined to be clearly shrinking loops (SADLs) is indicated in the table so that the SADs results can be compared with those of the SADLs.  (Both SADs and SADLs are clearly observed in the 2002 April 21 TRACE event.)  Also indicated is the position of the flare on the Sun.  Flares beyond the limb are given a limb designation for the instrument field of view (FOV).  Flares occurring on the disk (within $\sim50^{\circ}$ from disk center) yield unreliable trajectory information due to the inability to accurately measure heights above the surface; therefore, their results are treated as detections only and removed from the following statistical analysis.   

It should be noted that the number of flows being reported are those that were deemed to be the most reliable and complete although additional flows (\textgreater~50) have been processed.  Inevitably, only a portion of the flows could be tracked for most flares due to noise, image quality, cadence, etc. (particularly for SXT) -- hence the need to process many flares in order to build up a catalog of flow parameters.  The high resolution (0.5~arcsec/pix), high energy (GOES X1.5+) selected TRACE flares yield by far the highest number of clearly defined, easily-trackable SADs and SADLs, but even in those flares, there is substantial untrackable downflowing motion whose shrinkage energy contribution cannot be included in the final total estimates.  This untracked motion is primarily present during the impulsive phase of the flare.

\subsection{\label{sadsiisec:uncertainties}Sources of Uncertainty}

Several variables were measured for each flow including height, velocity, acceleration, area, magnetic flux, and shrinkage energy.  A description of these measurements can be found in \cite{mckenzie-savage_2009}~(Section~2 therein).  Flows that were either too difficult for the automatic routine to follow or contained shrinking loops were tracked manually instead.  Although conservatively determined by visually judging the cross-sectional diameters and extrapolating a circular area, areas assigned manually are typically smaller than those determined using the automatic threshold technique.  All of the manually evaluated flows are assigned a single area per flare whereas the automatically determined areas vary in time.  The manual trajectories are thus better determined though the flow sizes are not temporally flexible.  This is especially true with SXT data because of the low spatial resolution.  As noted in \cite{mckenzie-savage_2009}~(Section~3 therein), degrading the resolution of TRACE images to that of SXT's half-resolution leads to flow areas comparable to that of SXT.  The result is that several smaller flows become undetectable, some flows that are near one another spatially are combined, and several of the flow ``heads" are merged with their trailing ``tails" making them appear larger.  The square root of the largest area extent is used as the error on the flow positions.

An additional large source of error is the initial height location.  This height is biased by instrument-dependent detection capabilities (e.g. dynamic range and FOV) and is limited by the low emission high above the arcade.  The initial height detection limits the path length of the measured flow trajectory which is used in the shrinkage energy calculation.  Of larger consequence, the initial height determines the initial magnetic field invoked from the PFSS model (\citeauthor{schatten-wilcox-ness_1969}~\citeyear{schatten-wilcox-ness_1969}; \citeauthor{schrijver-derosa_2003}~\citeyear{schrijver-derosa_2003}) to calculate the flux and shrinkage energy.  The ``Cartwheel CME" flare from 2008 April 9 (\citeauthor{ko_2010}~\citeyear{ko_2010}; \citeauthor{landi_2010}~\citeyear{landi_2010}; \citeauthor{patsourakos-vourlidas_2010}~\citeyear{patsourakos-vourlidas_2010}; \citeauthor{savage_2010}~\citeyear{savage_2010}) is a nice example of being able to observe near the actual flow initiation region due to long exposure durations enabled by limb-obscured footpoints.  This is a rare example, however, because active regions are not often observed for long after crossing the western solar limb and it is difficult to anticipate flares prior to crossing the eastern limb.  AIA, an EUV imager aboard the recently-launched Solar Dynamics Observatory (SDO), will improve the number of these necessary limb observations since it observes the full solar disk continuously.

\begin{figure}[!ht] 
\begin{center}

\includegraphics[width=0.6\textwidth]{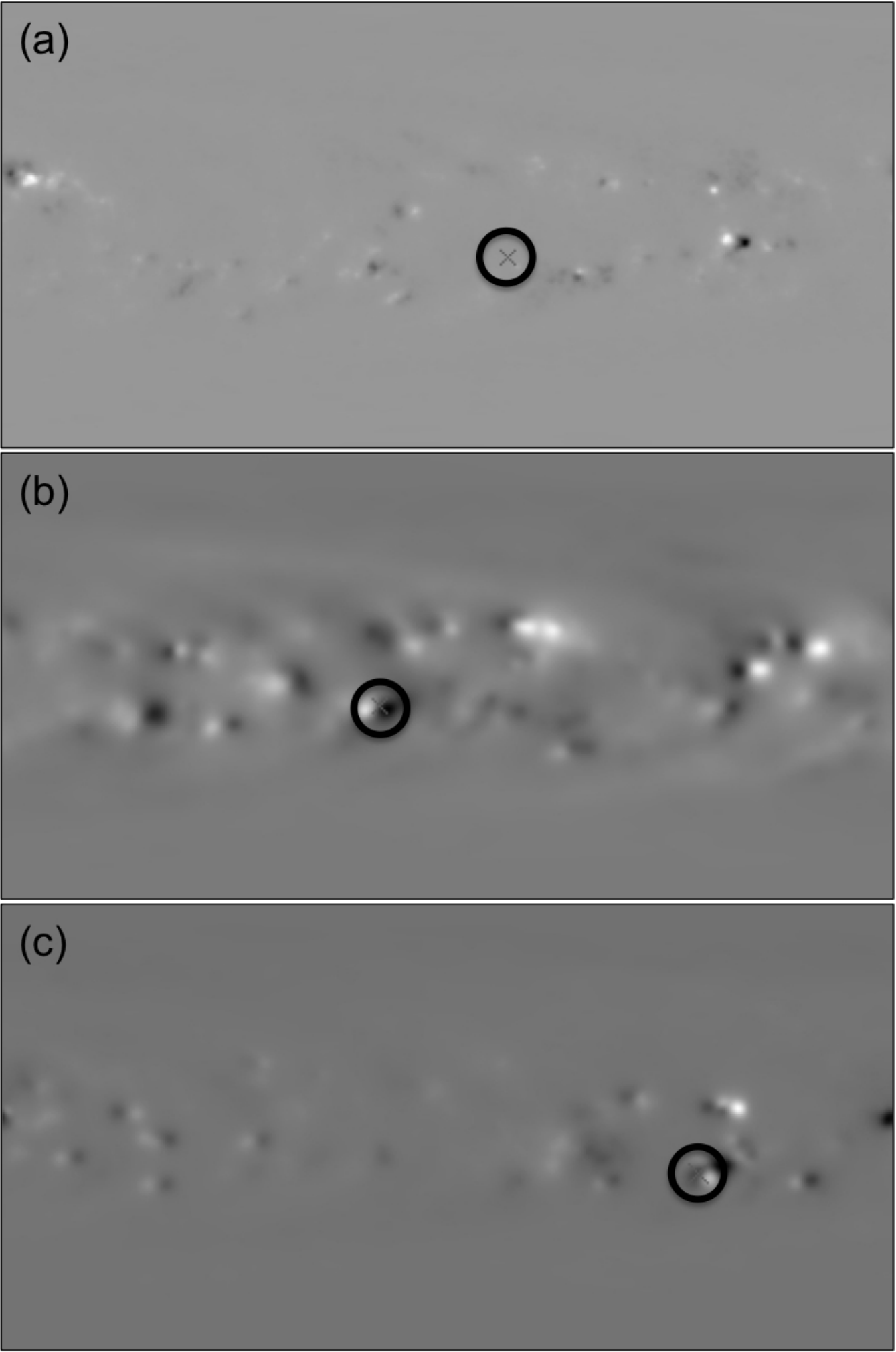}

\caption{Radial magnetic fields derived from PFSS magnetic field modeling for the active regions producing flares observed by TRACE on a) 2002 July 23, b) 2002 April 21, and c) 2003 November 4.  The initial height of the first flow per flare was used to create the figures.  The footpoint region is circled.  The east limb event in (a) predicts much weaker fields than the other two west limb events.}
\label{radial_field}
\end{center}
\end{figure}

Also noted in \cite{mckenzie-savage_2009}, the PFSS model itself is another source of uncertainty considering that the flows are associated with flaring active regions which are expected to be non-potential.  It was estimated that the uncertainties from the model do not exceed about 30\% (Priest, private communication).  The larger contributions to the magnetic field uncertainty are therefore the height input into the model, the footpoints assigned to the flows, and the flare's position on the Sun.  Limb flares provide the optimum viewing angle for flow detections; however, the magnetograms used to extrapolate the magnetic fields into the corona are more reliable on the disk.  Additionally, modeling for the east limb is even less reliable because magnetograms prior to the flare are unavailable.  This discrepancy is shown in Figure~\ref{radial_field} as the footpoint used for an X-class east limb flare (2002 July 23) is circled in panel (a) while the footpoints for X-class west limb flares (2002 April 21 \& 2003 November 4) are circled in the bottom two panels.  The initial height of the first flow per flare was used as a basis for the magnetic field represented in the panels.  There is some variance, however, in the initial magnetic field strength between flows because the field extrapolation depends on initial height.  Note the strong magnetic field indicated by the magnetograms for the west limb flares compared to that of the east.  Several of the flares in Table~\ref{sads_list_analyzed} occurred on the east limb (coordinates are shown in Table~\ref{sads_list_full}).  Magnetic fields are still estimated for these flares.  The effect of the underestimated magnetic field becomes apparent in Figure~\ref{quart3}~(c).

Velocity results may be biased by the inability to track flows that are so fast they only show up in one frame or so slow that they are unobservable in the difference images.  Most flows appear to be moving well within these constraints.  Flows with areas near the resolution of the instrument are also difficult to detect which contributes an additional bias to the statistics.  The obvious instrumental effect on area measurements is shown in Figure~\ref{quart2}~(a).

\clearpage

\section{\label{sadsiisec:analysis}Analysis}

Considering these substantial uncertainty sources, flow measurements should be taken as imprecise; however, the large number of fairly well-defined flows (total of 387) tracked in the flares listed in Table~\ref{sads_list_analyzed} make it possible to consider ranges and trends in the data.  Only the limb flares are considered hereafter which reduces the total number of flows to 372.  The upflows (moving away from the solar surface) tracked and analyzed for XRT 2008~April~9 (see \cite{savage_2010}, Section~3.2 therein) were also excluded bringing the total number of flows to be considered in the following sections to 369. 

Note that all positional information provided in the following section is based upon the centroid of a trough location.  Theoretically, tracking the head of the trough would constitute a more accurate assessment of its position because the centroid position may include portions of the growing wake behind the actual shrinking loop which could cause the velocities to appear slightly smaller.  However, the automated routine relies on thresholds to determine the areas, and since each image is at least slightly different -- especially when the exposure durations vary -- the heads become much less reliable positions than the centroids.  Because the areas are all the same for the manual routine, the centroid and head positions are merely shifted from one another.

\subsection{\label{sadsiisec:quartiles}Synthesis of Frequency Diagrams}

The following plots synthesize the flow measurement results from all of the flares.   Each plot shown in Figures~\ref{quart1},~\ref{quart2}, and~\ref{quart3} consists of a quartile plot in the left panel and a histogram in the right.  For the quartile plots, the measurement is plotted against the instrument (or instrument combination) being considered (S: SXT; X: XRT; T: TRACE; L: LASCO; All: S\&X\&T\&L).  For Figures~\ref{quart1} and~\ref{quart2}, the left [red or purple] box-and-whisker range per instrument (or instrument combination) represents SADs measurements while the right [blue or purple] one represents SADLs measurements.   For Figure~\ref{quart3}, the east [pink or green] and west [olive or green] limbs are compared instead of SADs to SADLs.  The lines (or whiskers) extending from the boxes indicate the full range of the data.  The boxes span the range of the middle 50\% of the data.  The [white] line through the box indicates the median of the data.  Along the top of these plots, the number of flows used to derive the associated measurements is labeled.  The combination of the data in the final two [purple or green] box-and-whisker plots is contained within the histogram panel.  The median of the histogram is displayed in the legend.

LASCO measurements are not included in Figures~\ref{quart2} or~\ref{quart3} since its resolution (11.4~arcsec/pix) is so much poorer and its observational regime high above the limb ($>$~2.5~R$_{\odot}$ for C2) is so very different from that of the other instruments, making comparisons more complicated.  Other issues with LASCO comparisons include:  1) Deriving magnetic fields at such heights is not applicable with our method (i.e. the LASCO heights are typically beyond the limit for reasonable extrapolations from the PFSS model) and 2)  Determining precise footpoints without coincidental data from other instruments is nearly impossible.  The total number of flows under consideration after removing those observed by LASCO is 358.

\subsubsection{Velocity \& Acceleration}

There is general agreement between SADs and SADLs, the instruments, and the SXR versus EUV bandpasses for the average velocity, initial velocity, and acceleration measurements (Figure~\ref{quart1}).  Note that the initial velocity and acceleration plots do not incorporate all 369 available flows.  Instead, only those flows tracked in at least 5 frames were included because these measurements rely on fitting the trajectories to a 2D polynomial fit.  Using fewer than 5 points leads to unreliable results.  Also note that a positive downflow acceleration means that the flow is slowing.  

\subsubsection{Area}

A strong correspondence between instrument resolution (SXT: 2.5~-~4.9~arcsec/pix; XRT: 1~arcsec/pix; TRACE: 0.5~arcsec/pix) and measured area is shown in the initial area quartile plot (Figure~\ref{quart2} (a)).  The SADLs and XRT SADs area measurements are very strongly peaked due to their manual assignments -- hence the lack of distinct quartiles.

\subsubsection{Height}

The initial height ranges (Figure~\ref{quart2} (b)) show decent agreement between the ranges; however, there is a fair amount of scatter in the medians which requires more detailed understanding of the analyzed flares to explain.  The initial heights for both SADs and SADLs observed by SXT offer very good agreement.  XRT observations, while agreeing with SXT's range of initial heights, show no agreement between SADs and SADLs.  This discrepancy is due to a combination of factors:  1) XRT has observed very few SADs near the limb as yet.  2) XRT observations are rarely sufficiently exposed to illuminate the supra-arcade region; therefore, XRT SADs have only been observed nearer to the solar surface.  3) The SADLs observed by XRT are derived from the ``Cartwheel CME" flare \citep{savage_2010} during which the footpoints were obscured by the limb enabling very long exposure durations.  In fact, a disconnection event associated with this flare (\cite{savage_2010}, Section~3.2 therein) was observed at nearly 190 Mm above the solar surface, which is at the max of the combined instrument ranges.  TRACE's temperature coverage targets plasma on order of 1 MK with some overlap in the 11~-~26~MK range with the SXR imagers.  The image exposure durations are also optimized to observe the flaring region near the solar surface. Consequently, the observed initial heights of SADs and SADLs measured with TRACE are limited to the region near the top of the growing post-eruption arcade where the hot plasma in the current sheet is most illuminated.  This results in initial heights lower than many of those reported for SXT and XRT.

The change in heights shown in Figure~\ref{quart2} (c) are naturally flare and FOV dependent.  Even so, there is general agreement between SADs, SADLs, and instrument except for XRT.  The explanation for this XRT discrepancy is the same as that for the initial height XRT discrepancy described above (i.e. the flows for the ``Cartwheel CME" flare could be tracked further through the FOV).

\subsubsection{Magnetic Measurements}

Figure~\ref{quart3} (a) is provided as a visual reference for the initial magnetic fields which are used to calculate the magnetic flux ($\Phi = B \times A$, Figure~\ref{quart3} (b)) and the shrinkage energy ($\Delta W = B^2 A \Delta L / 8\pi$, Figure~\ref{quart3} (c)).  (Refer to \cite{mckenzie-savage_2009}, Section~4.5, for a detailed description of the shrinkage energy calculations.)  The initial magnetic field estimates for the TRACE flares are larger than the majority of those from SXT and XRT possibly because 1) the TRACE flares analyzed are highly energetic according to their GOES classifications and 2) the flows are observed closer to the surface (see Figure~\ref{quart2} (b)) where the magnetic field is stronger according to the PFSS model (\citeauthor{schatten-wilcox-ness_1969}~\citeyear{schatten-wilcox-ness_1969}; \citeauthor{schrijver-derosa_2003}~\citeyear{schrijver-derosa_2003}).  East and west limb measurements are compared in Figure~\ref{quart3} to show the effect of using less reliable east-limb magnetograms.   The tendency for west limb flares to have stronger initial magnetic field estimates is noticeable in Figure~\ref{quart3} (a) and carries through into the initial magnetic flux and shrinkage energy plots (b \& c, respectively).  The dichotomy is most noticeable in the shrinkage energy estimates due to the $B^2$ component.    

\begin{figure}[!ht] 
\begin{center}

\includegraphics[width=0.8\textwidth]{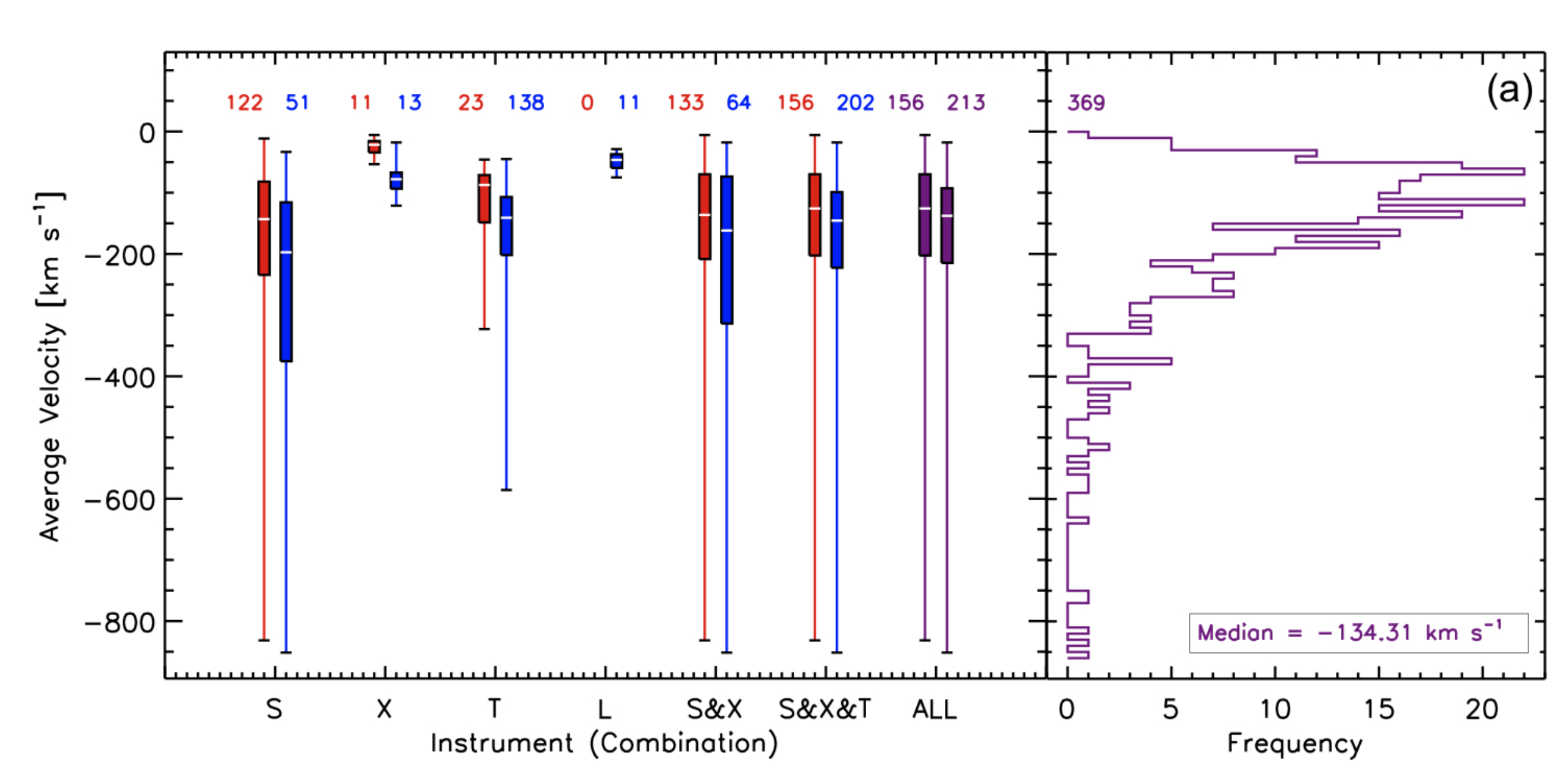}
\includegraphics[width=0.8\textwidth]{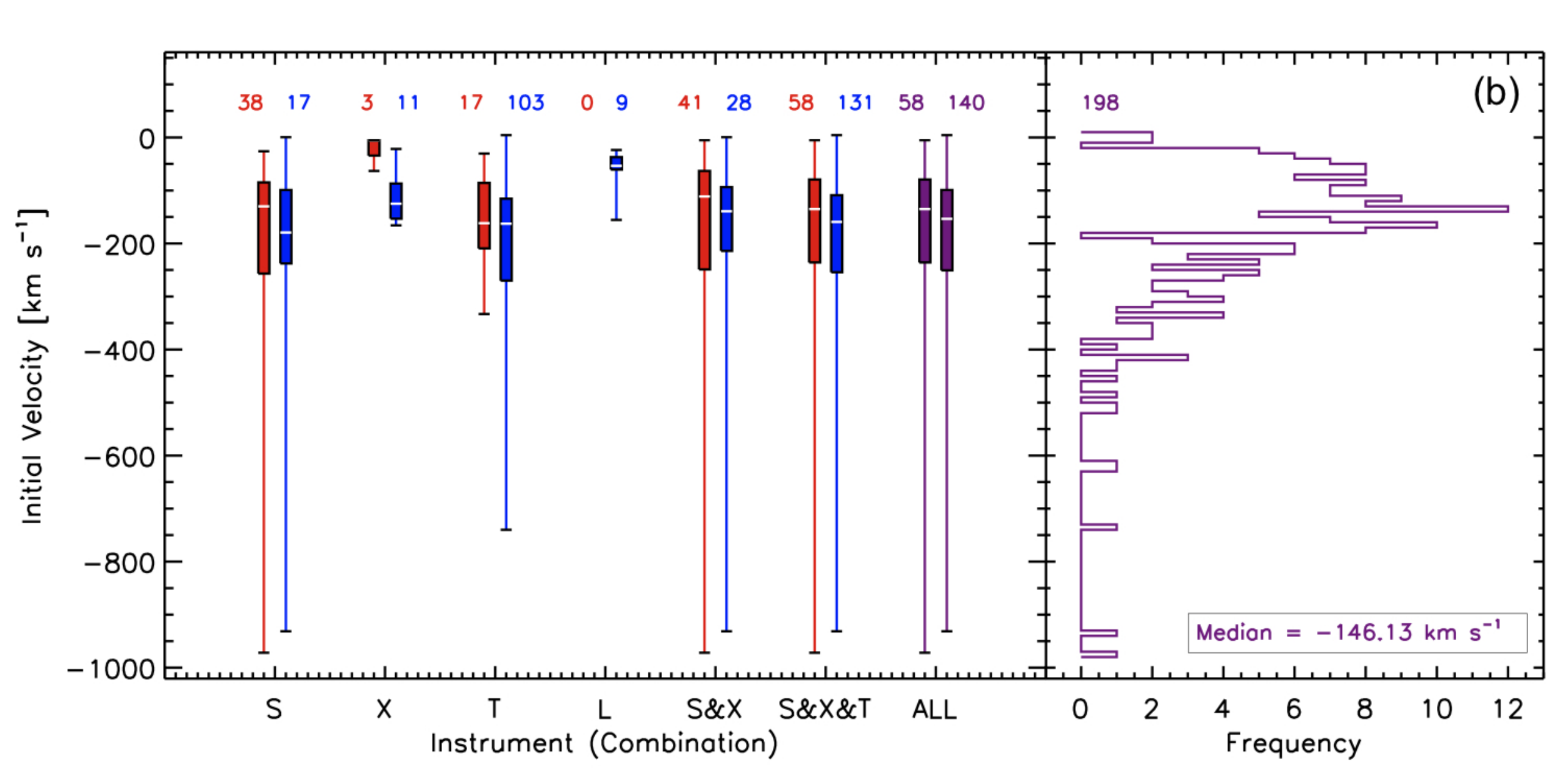}
\includegraphics[width=0.8\textwidth]{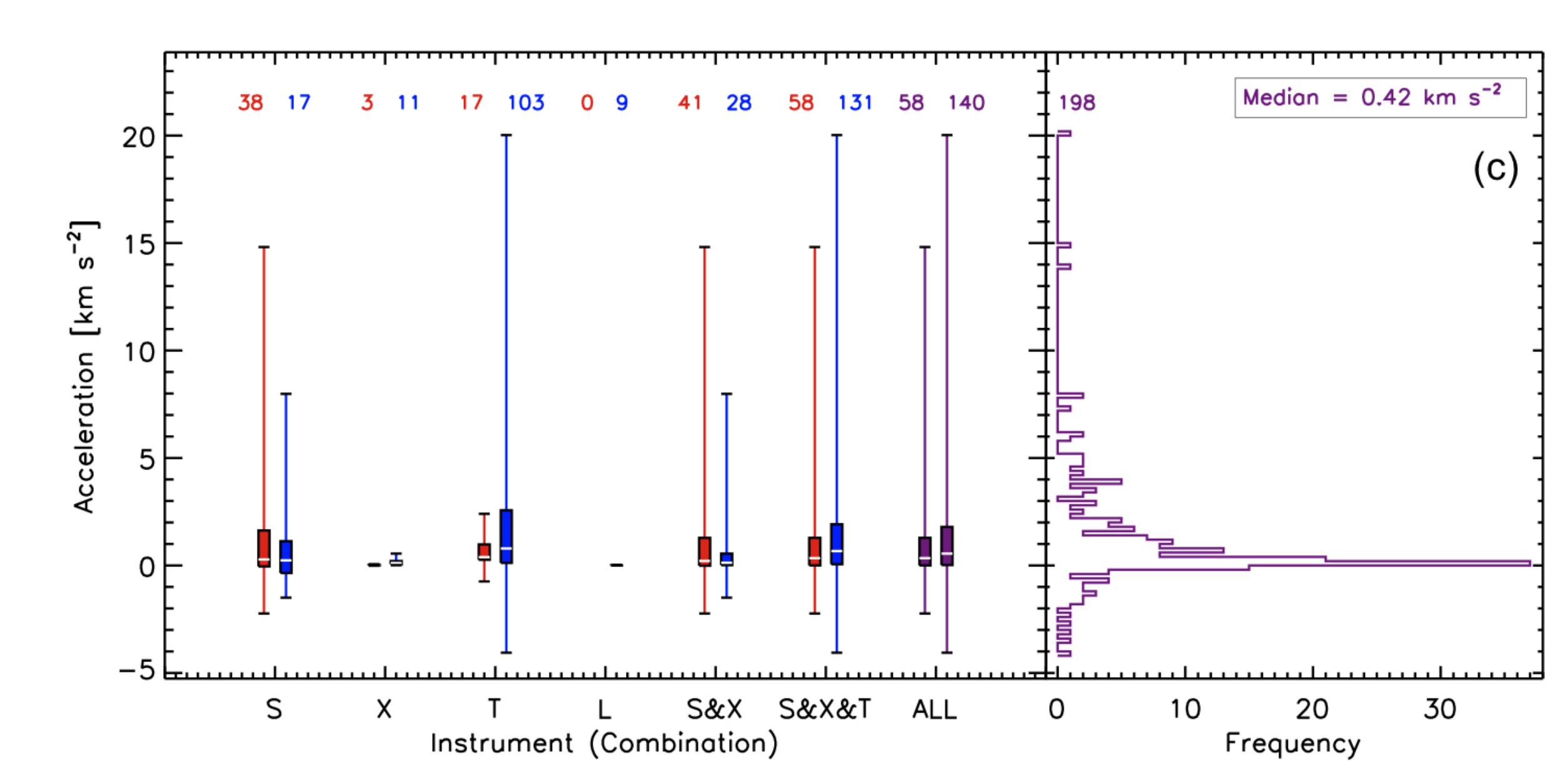}

\caption{Synthesis of the de-projected trajectory parameter estimates:  a)  De-projected average velocity.  b)  De-projected initial velocity.  c)  De-projected acceleration.  (Refer to the text for a detailed description of these figures.)}
\label{quart1}
\end{center}
\end{figure}

\begin{figure}[!ht] 
\begin{center}

\includegraphics[width=0.8\textwidth]{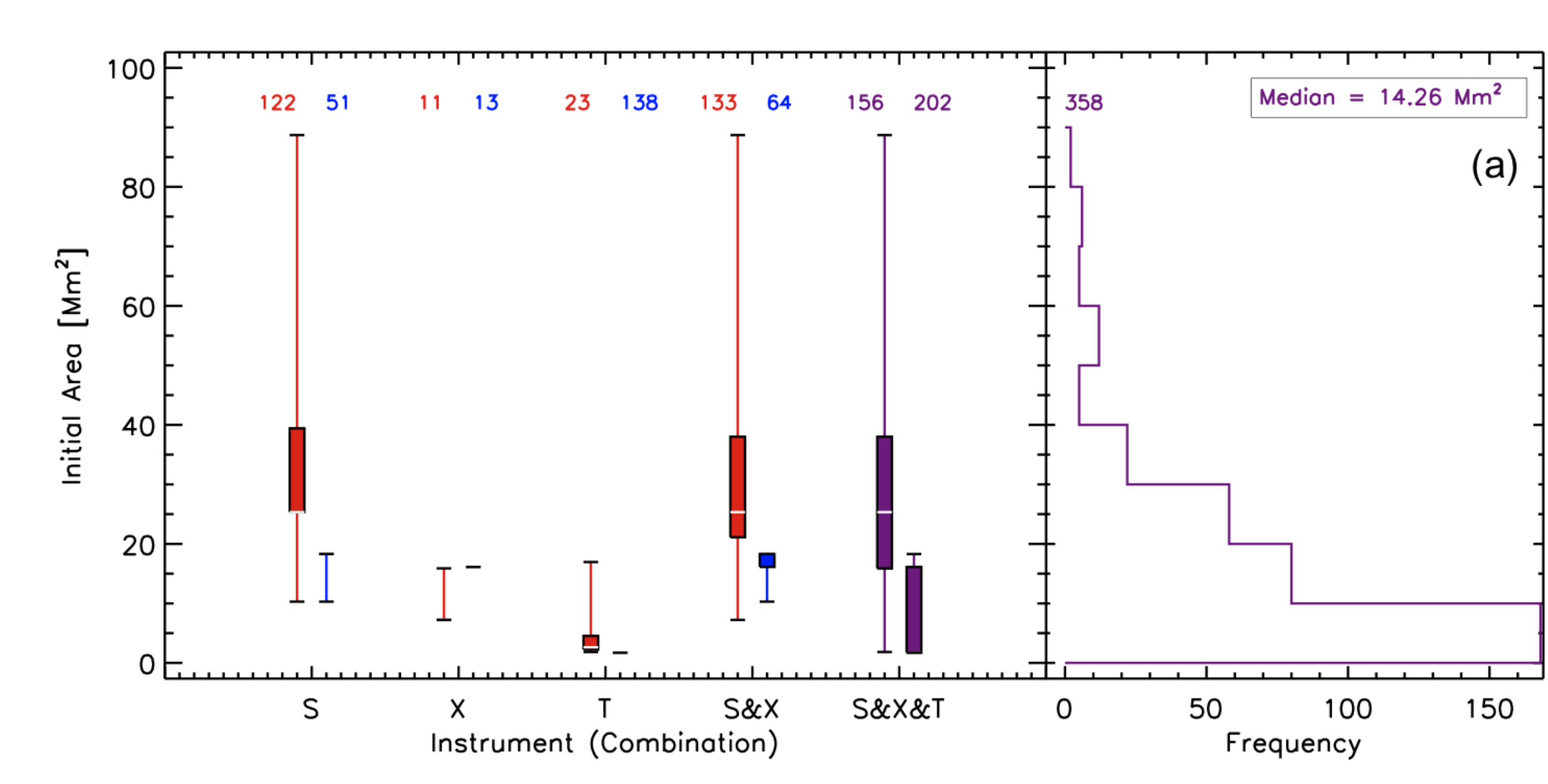}
\includegraphics[width=0.8\textwidth]{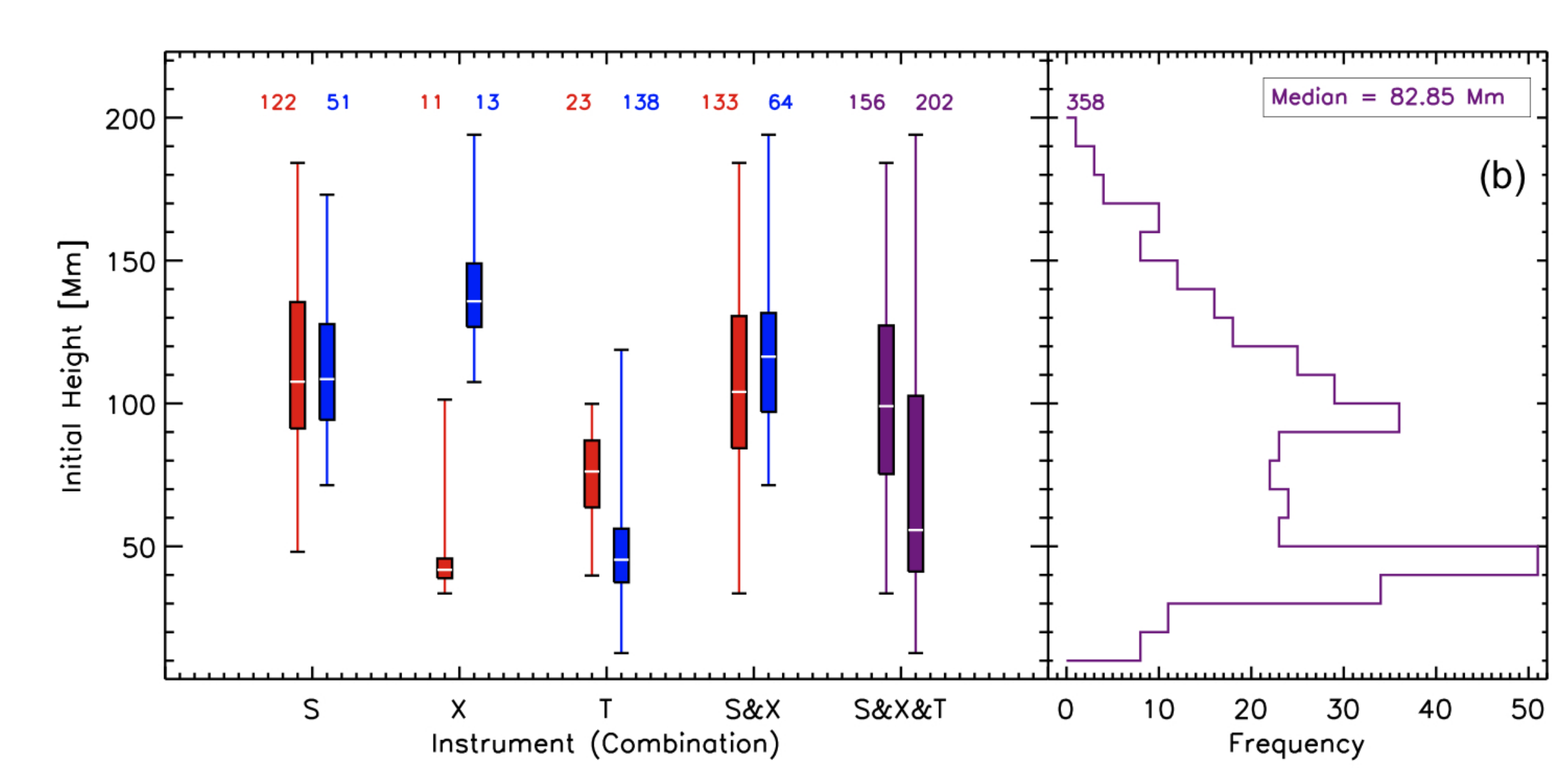}
\includegraphics[width=0.8\textwidth]{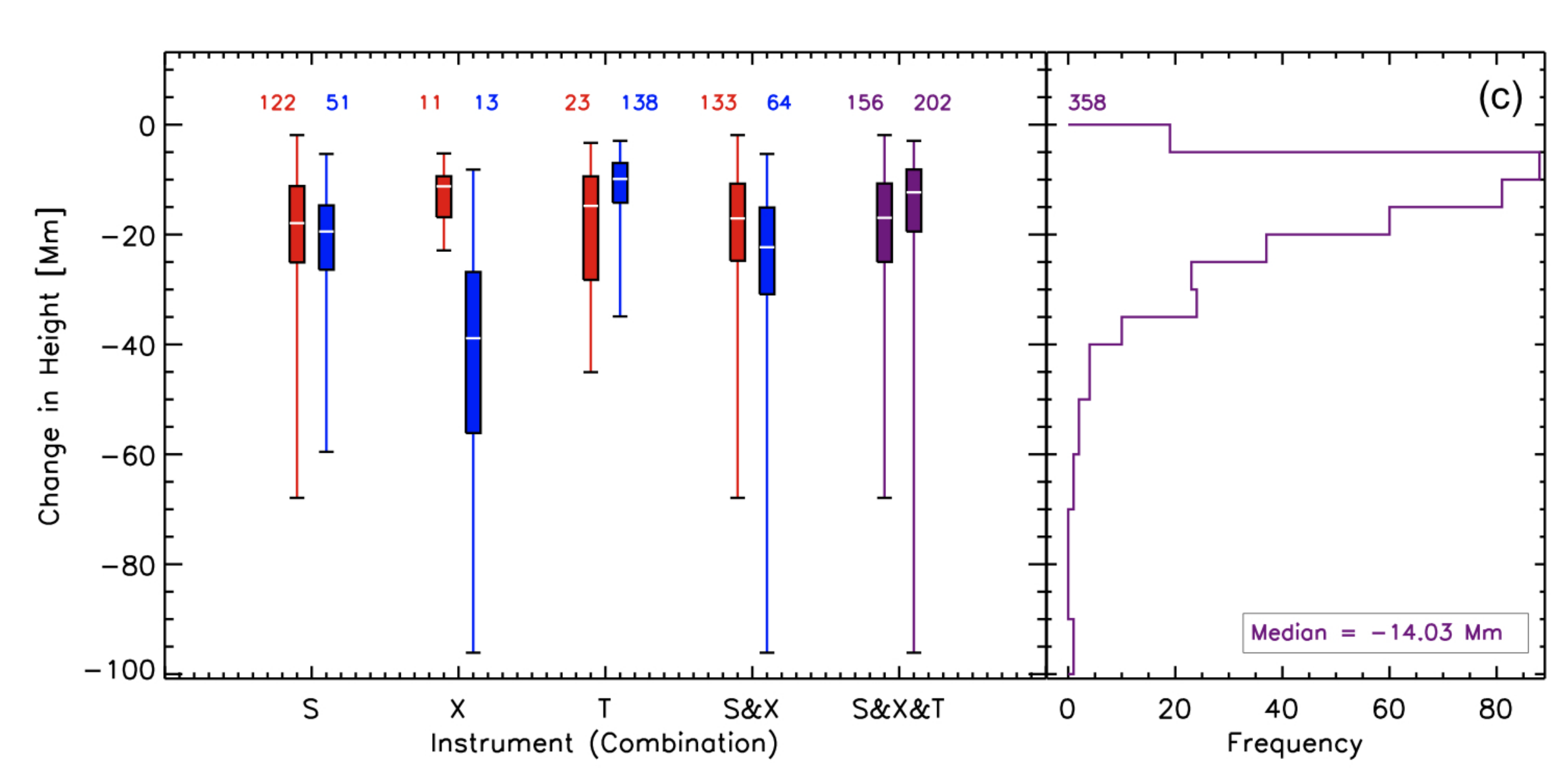}

\caption{Synthesis of the area and height parameter estimates:  a)  Initial area.  b)  De-projected initial height.  c)  De-projected change in height.  (Refer to the text for a detailed description of these figures.)}
\label{quart2}
\end{center}
\end{figure}

\begin{figure}[!ht] 
\begin{center}

\includegraphics[width=0.8\textwidth]{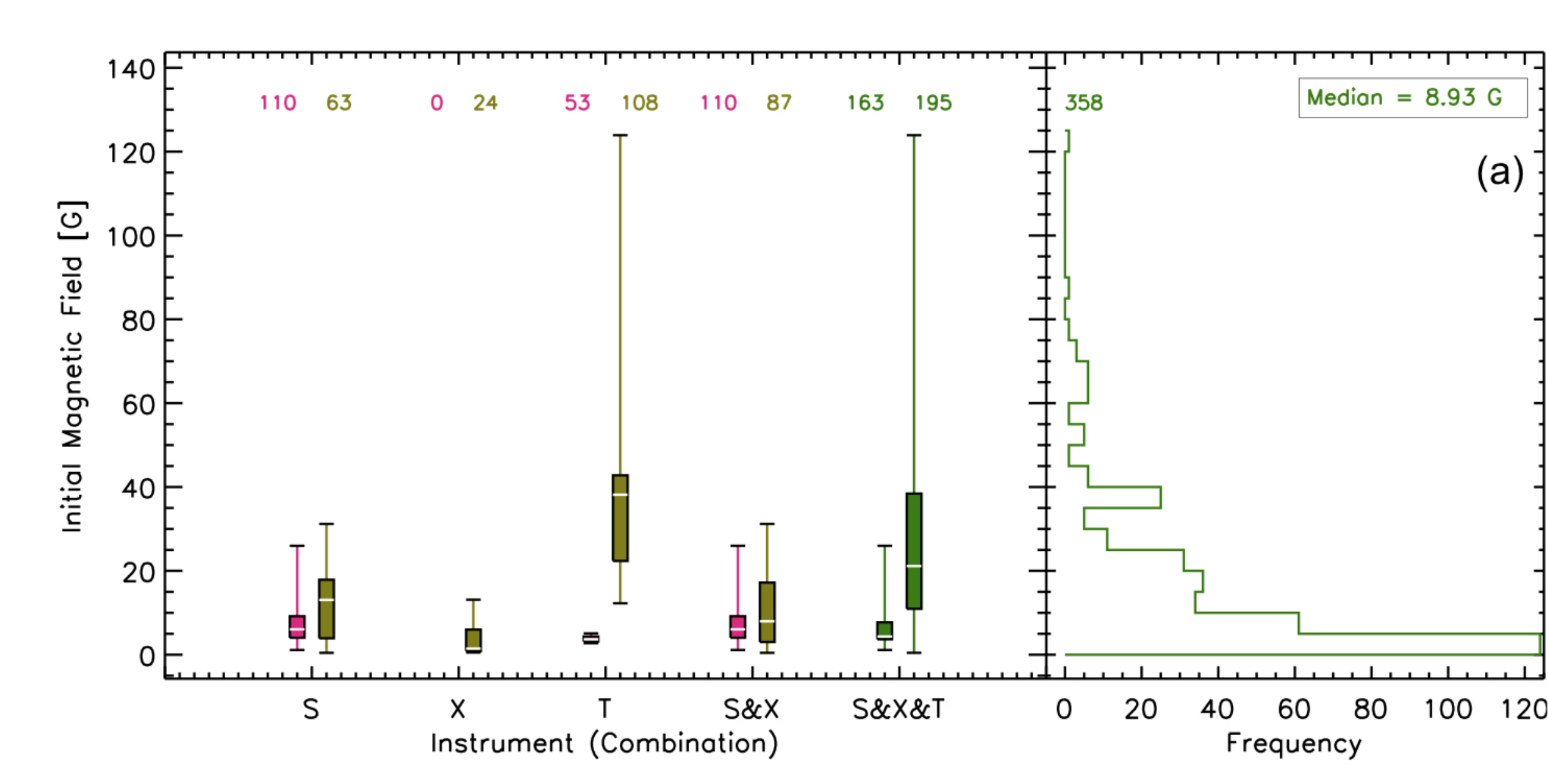}
\includegraphics[width=0.8\textwidth]{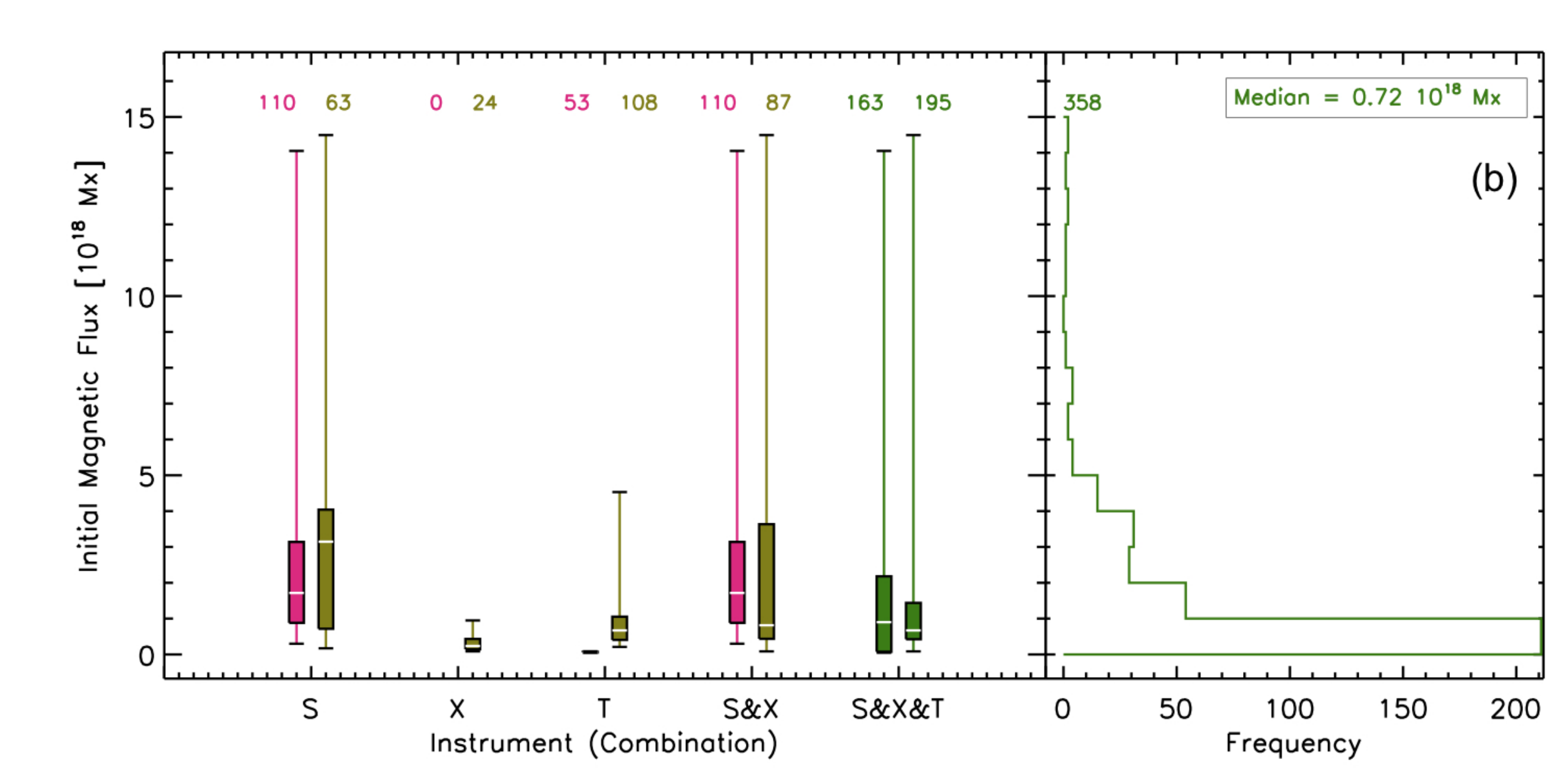}
\includegraphics[width=0.8\textwidth]{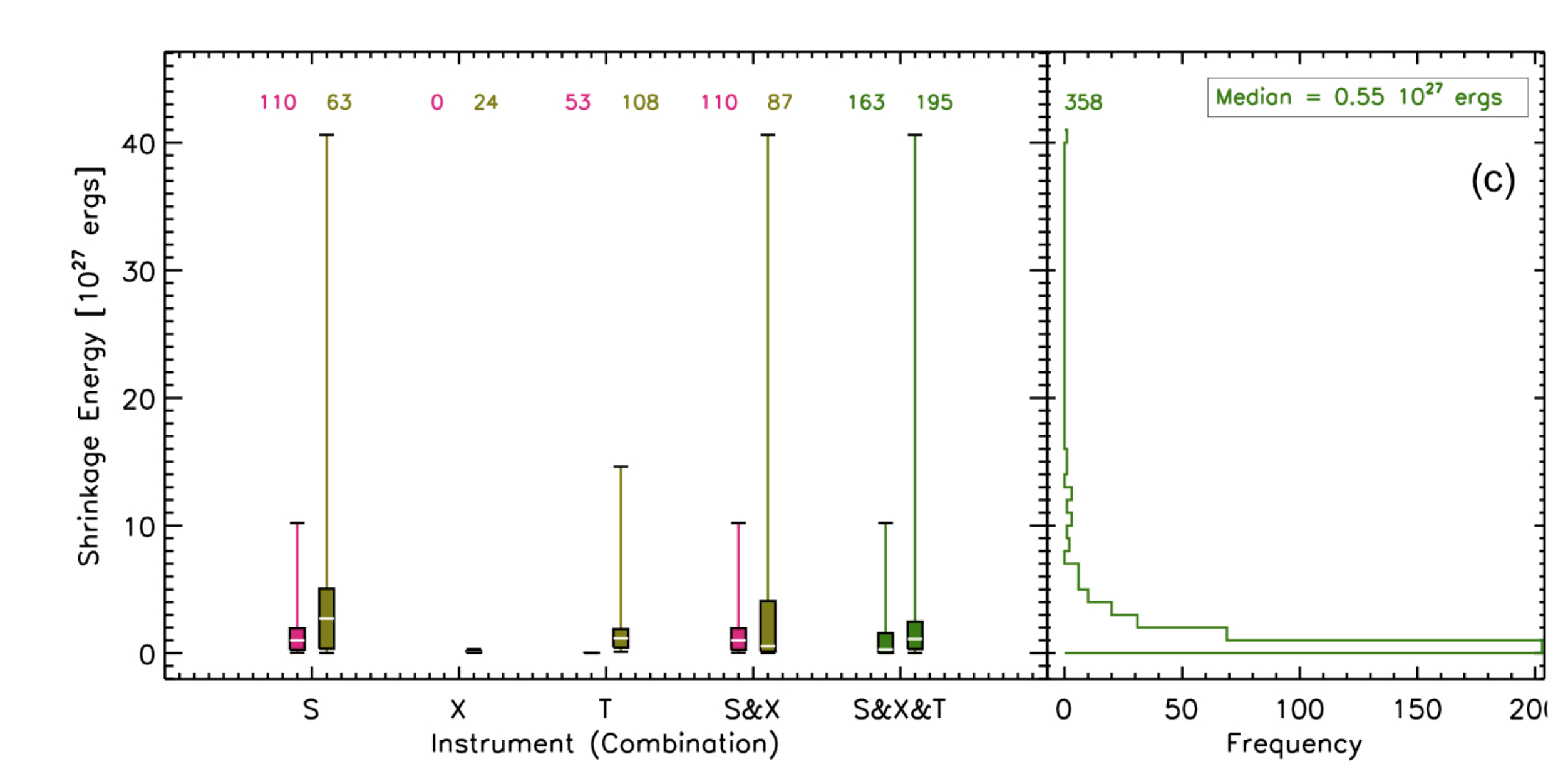}

\caption{Synthesis of the magnetic parameter estimates:  a)  Initial magnetic field.  b)  Initial magnetic flux.  c)  Shrinkage energy.  (Refer to the text for a detailed description of these figures.)}
\label{quart3}
\end{center}
\end{figure}

\clearpage

\subsection{\label{sadsiisec:trends}Trends}

\begin{figure}[!ht] 
\begin{center}

\includegraphics[width=0.49\textwidth]{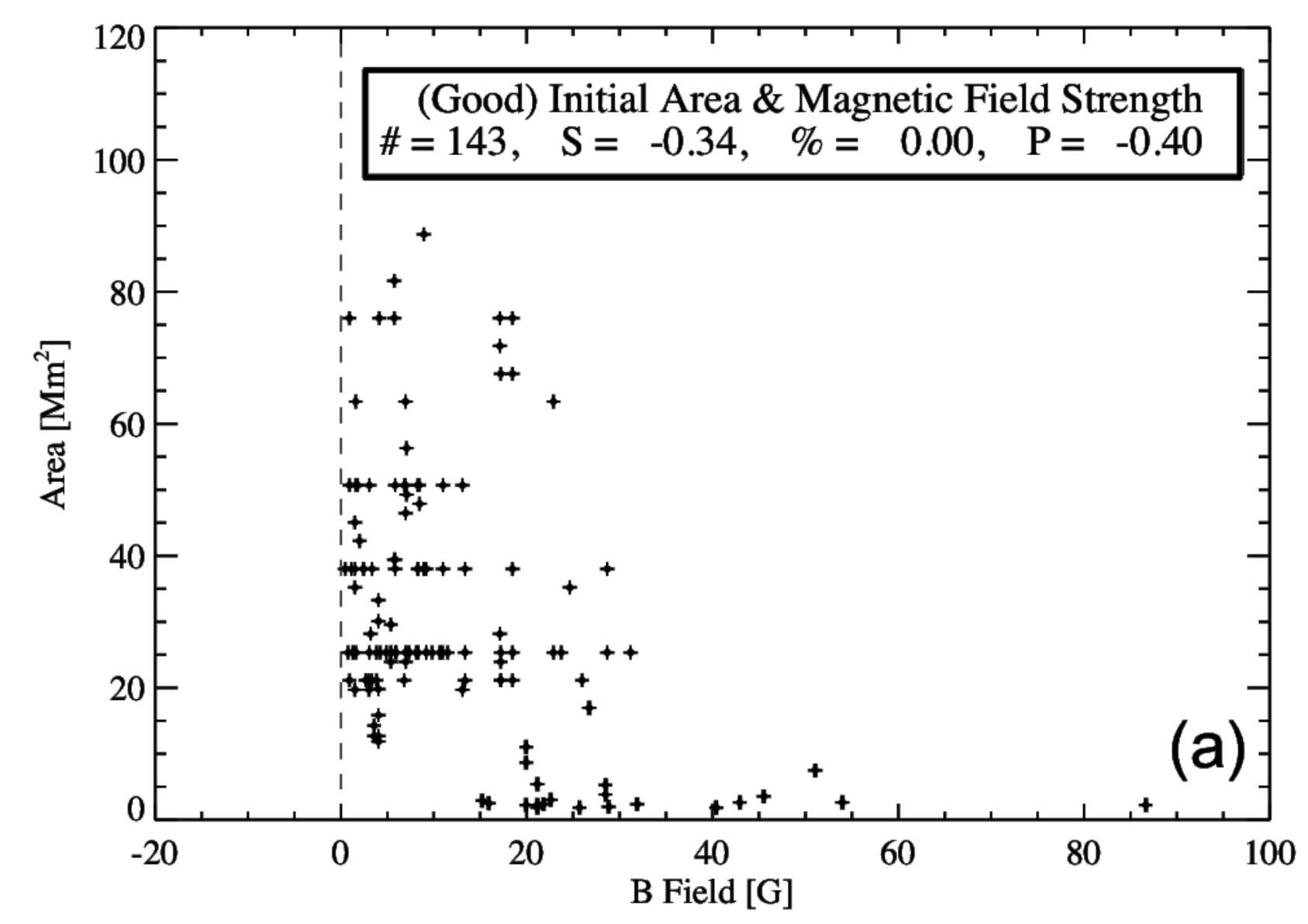}
\includegraphics[width=0.49\textwidth]{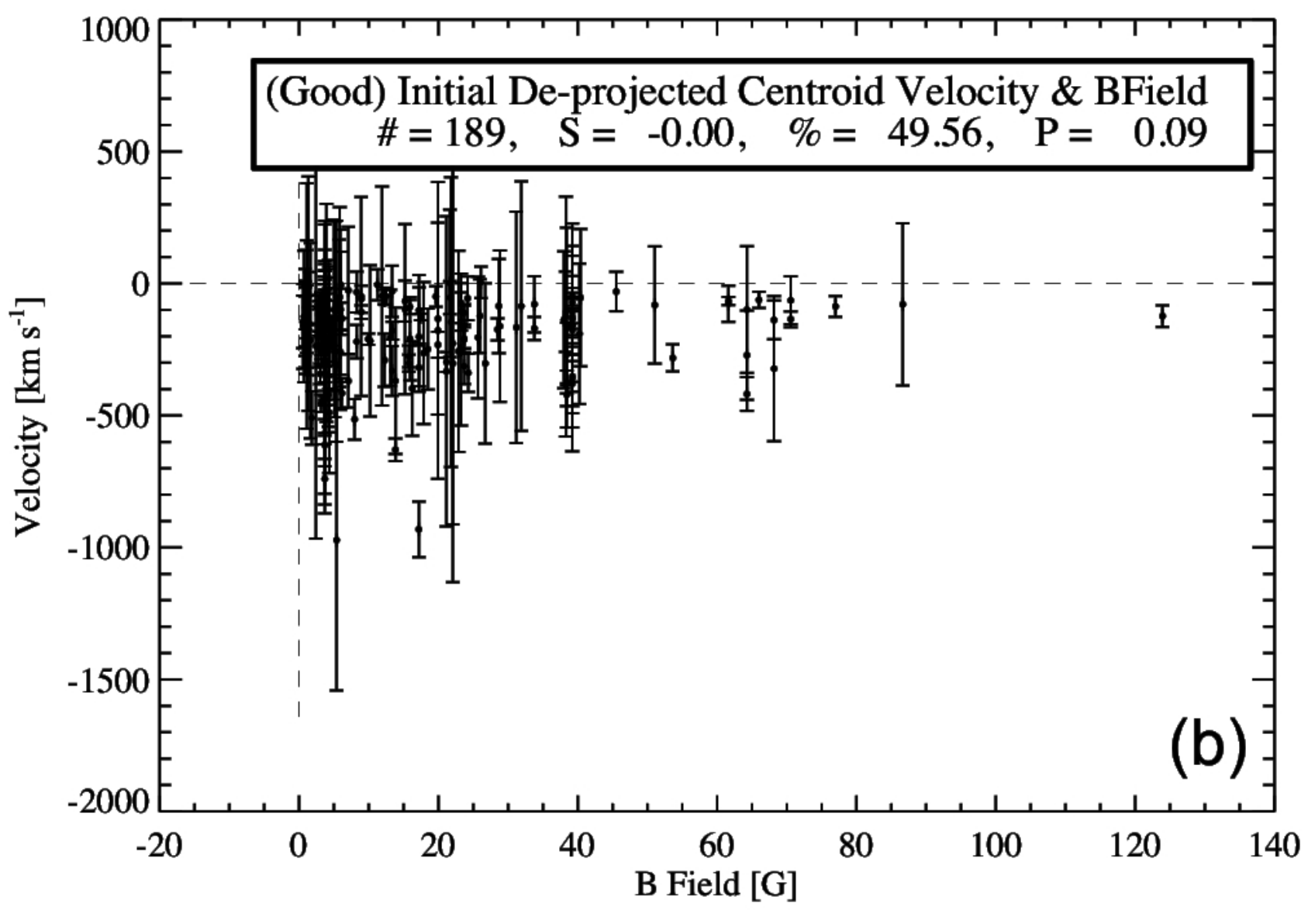}
\includegraphics[width=0.49\textwidth]{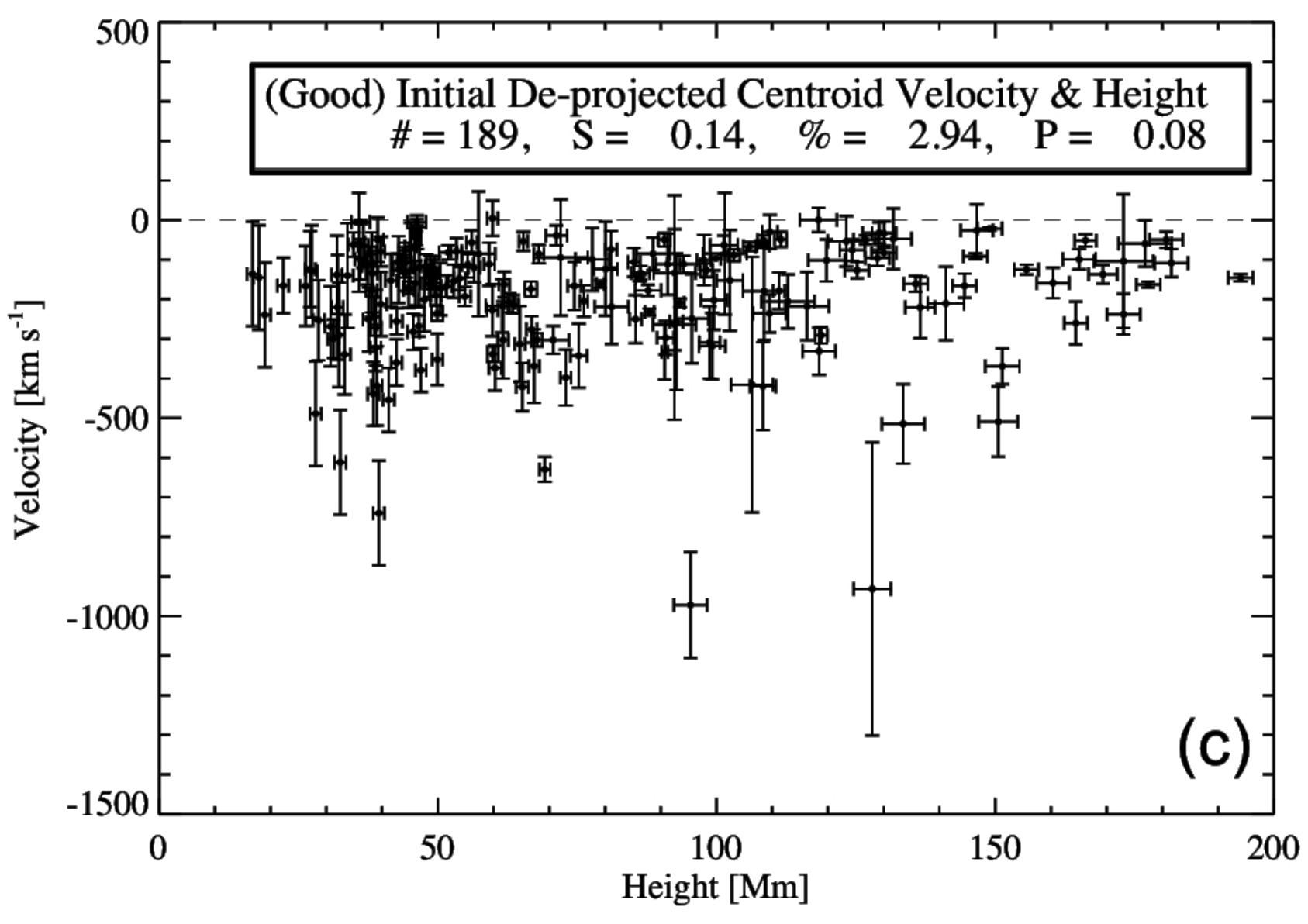}
\includegraphics[width=0.49\textwidth]{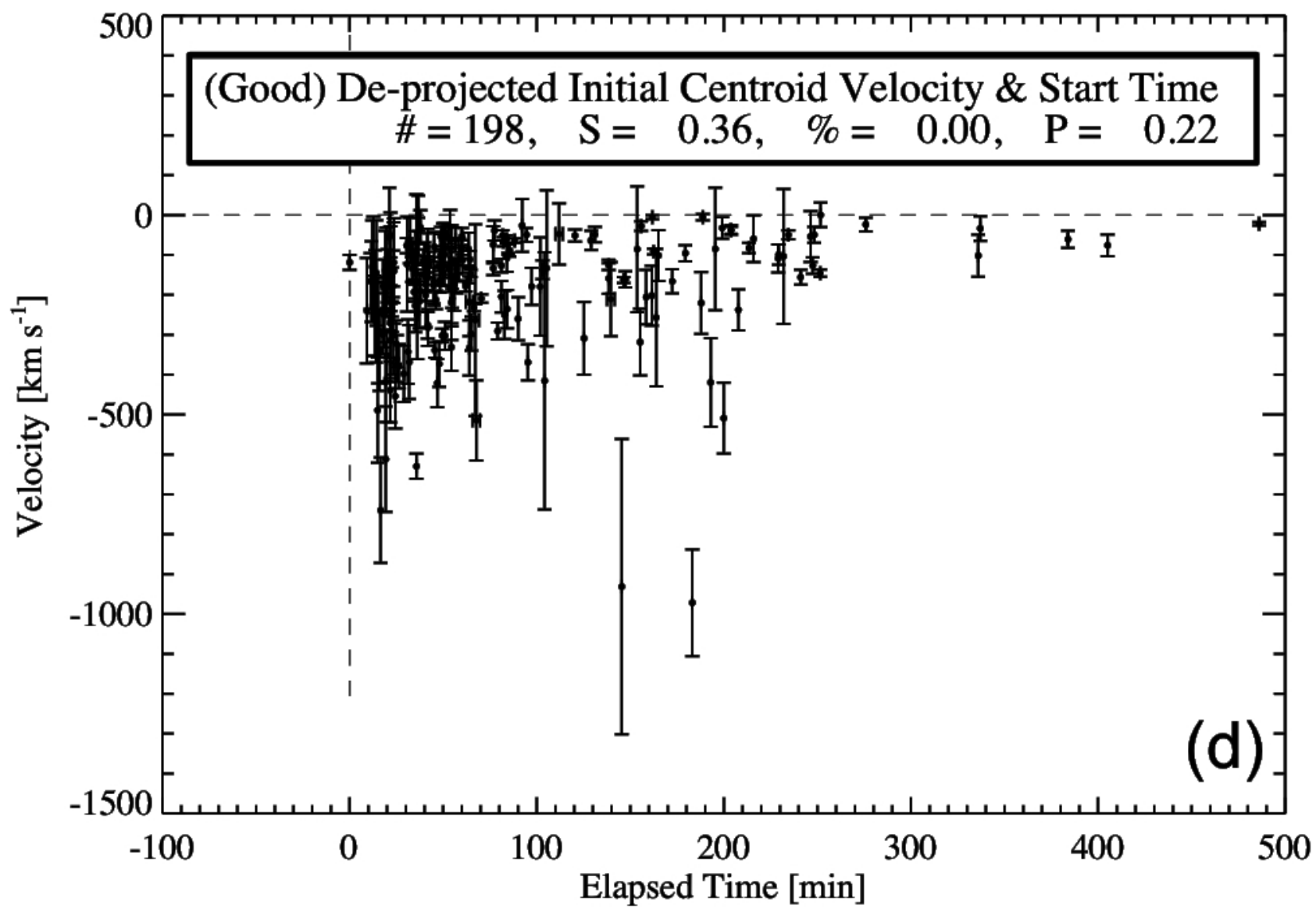}
\includegraphics[width=0.49\textwidth]{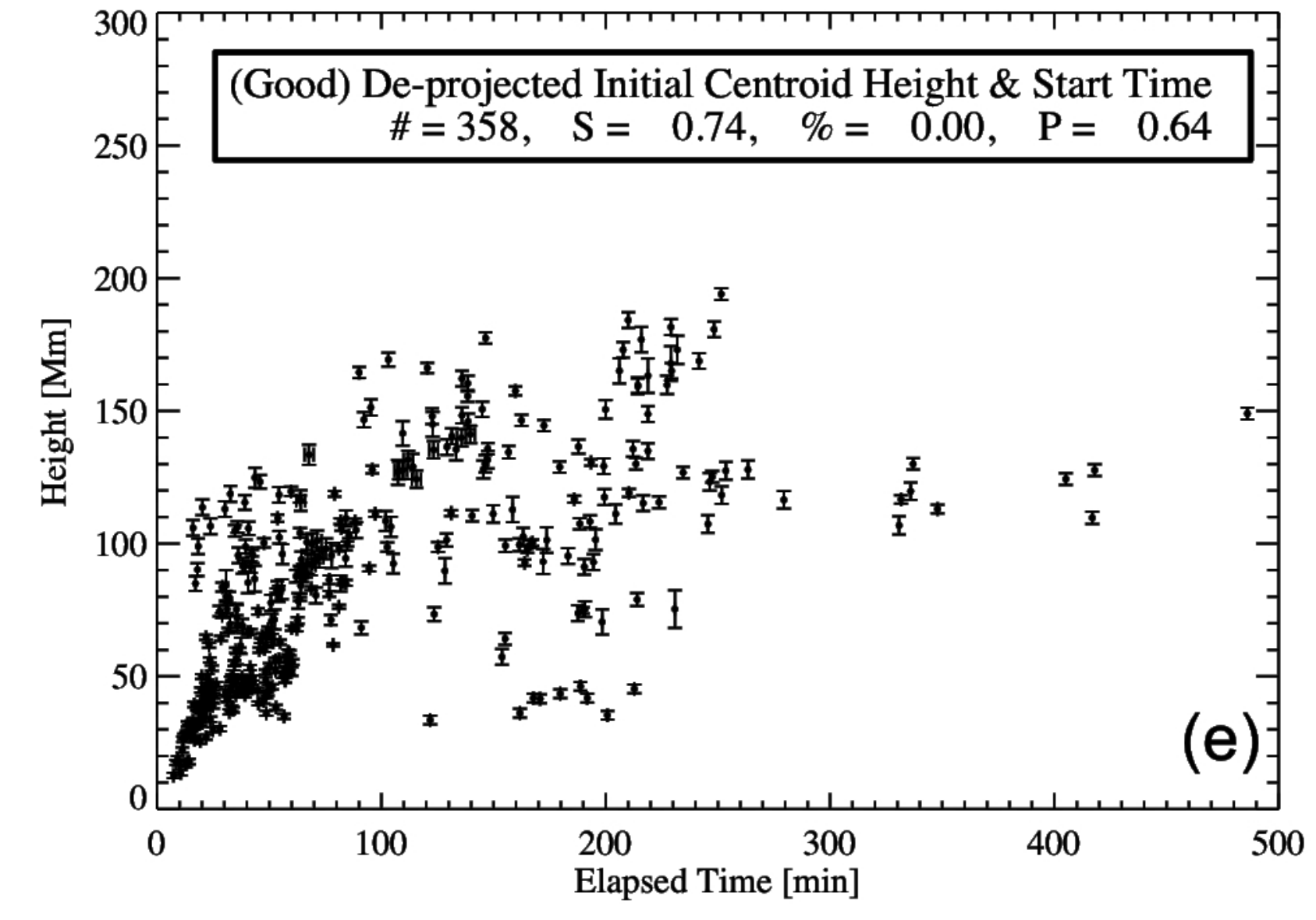}
\includegraphics[width=0.49\textwidth]{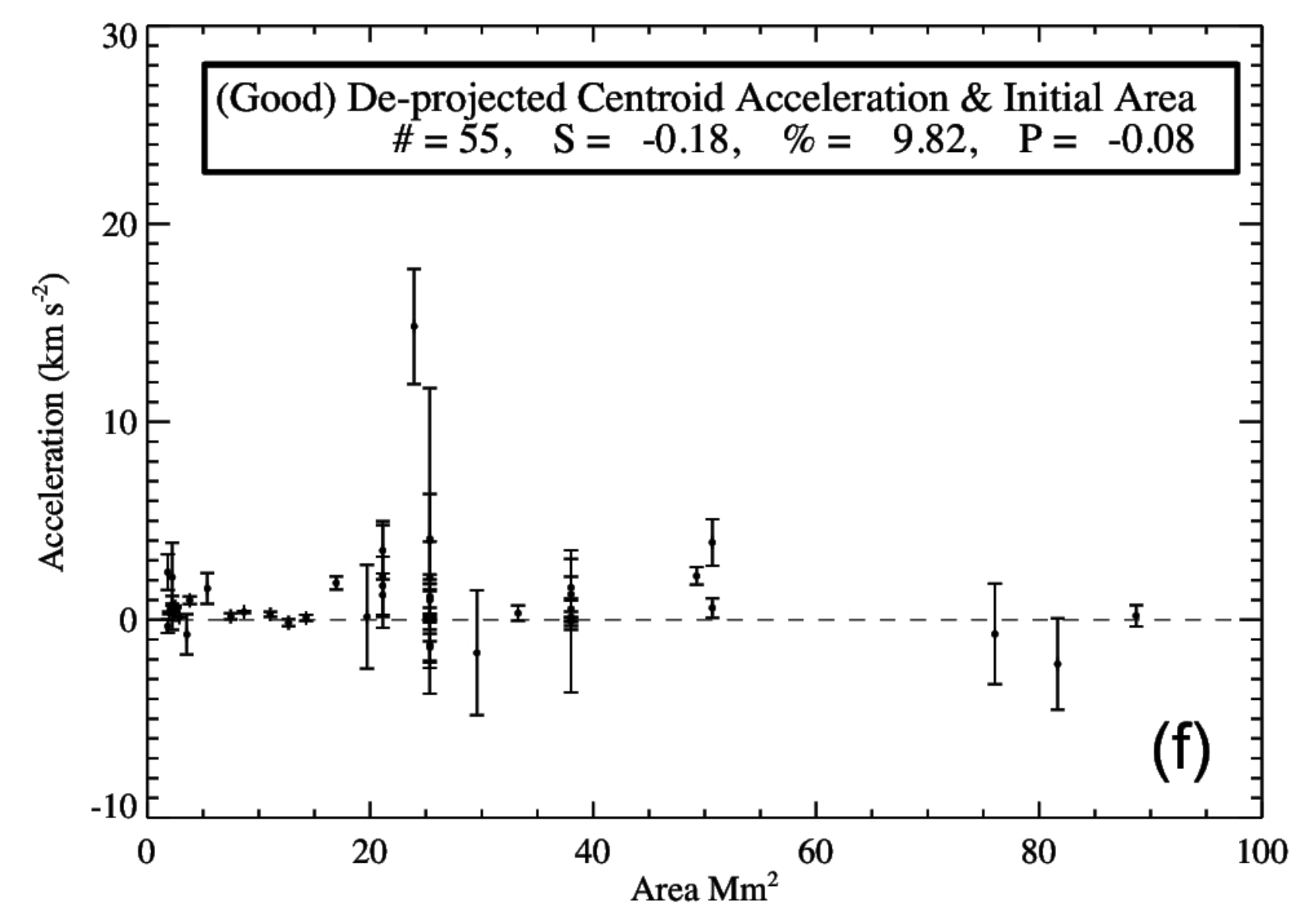}

\caption{Selected trend plots.  (a) Initial area (only for those measured via the threshold technique) versus initial magnetic field strength.  (b) Initial de-projected velocity versus initial magnetic field strength.  (c)  Initial de-projected velocity versus initial de-projected height.  (d) Initial de-projected velocity versus elapsed time.  (e) Initial de-projected height versus elapsed time.  The elapsed time is the time from the flare start time as indicated by the GOES light curves.  Contained with the legends are the number of flows used to create the plot (\#), the Spearman rank order correlation coefficient (S), the Spearman chance probability (\%), and the Pearson linear correlation coefficient (P).}

\label{trend_limb_selected}
\end{center}
\end{figure}

Figure~\ref{trend_limb_selected} presents some of the interesting trends from analysis of the downflow data.   Contained within the legends are the number of flows used to create the plot (\#), the Spearman rank order correlation coefficient (S), the Spearman chance probability (\%), and the Pearson linear correlation coefficient (P) \citep{hogg-craig_1995}.  The coefficients can range from 0 to 1 with 1 indicating a perfect correlation.  The Pearson coefficient applies to linear trends.  Negative coefficients indicate anti-correlations between the parameters.  The chance probability quantifies the likelihood of the data order occurring randomly.  LASCO flows are only included in Figure~\ref{trend_limb_selected} (d).  The number of flows used in panel (a) is based only on those measured with the automatic threshold technique.  Panels (b), (c), and (d) only include flows detected in at least 5 frames.

The apparent inverse relationship between field strength and flow area shown in Figure~\ref{trend_limb_selected}~(a) may be an artifact due to the larger detected areas at higher heights (instrument independent).  It is unclear as to whether this is a real effect or a result of poor signal to noise at such heights.  The lack of directly measurable magnetic field strengths may also be skewing the data set.  If the trend from Figure~\ref{trend_limb_selected}~(a) is real, then it suggests that a lower magnetic field strength results in a longer diffusion time (i.e. the amount of time that flux is able to reconnect) thereby allowing the flux tubes to increase in size.   

Figure~\ref{trend_limb_selected} (b) shows a very slight visual trend in favor of higher velocities with larger magnetic field strengths although the correlation coefficients do not support a trend.  Few velocity measurements are available for very large magnetic fields.  The lack of a strong trend is not so surprising, however, considering that the higher field strengths are associated with low heights, and these low initial heights are generally associated with flows near the top of the arcade where they are slowing considerably.  More precise initial positions would be required to verify any link between speed and field strength.

Plotting the initial velocity against the initial height (Figure~\ref{trend_limb_selected} (c)) also reveals a lack of correlation.  These parameters, however, are highly sensitive to precise initial positions.  There does appear to be a drop in speed at heights above 150 Mm.  The disconnection event from the ``Cartwheel CME" flare (\cite{savage_2010}, Section~3.2 therein) occurred near 190 Mm.  It is possible that there is a region of flow acceleration as it exits the reconnection site and before it slows as it reaches the arcade.  This plot may be indicating such a case; however, there are far too few high velocity points for this idea to be beyond speculative and one would expect for an acceleration region to be flare dependent.  Also, in the 2002 April 21 TRACE event, fast-flowing loops are seen to be descending early during the impulsive phase, yet they are not positioned favorably for proper tracking.  Knowing their precise speeds and initial heights could have a substantial effect on this plot in particular.

If approximating the flows as thin reconnected flux tubes, their velocity is expected to reflect the Alfv\'{e}n speed in the corona \citep{linton-longcope_2006} which has often been roughly estimated to be 1000~km~s$^{-1}$.  The flows are traveling nearly an order of magnitude slower than this estimate.  This discrepancy may partly be explained by the fact that the flow trajectories are incomplete and are being observed as they are approaching the post-eruption arcade.  Interestingly though, the downflow portion of the disconnection event from the ``Cartwheel CME" flare is estimated to have a maximum velocity of only about 150~km~s$^{-1}$.   (The upflow portion moves at about 120~km~s$^{-1}$.)

Figure~\ref{trend_limb_selected} (d) indicates that initial flow speeds are strongly correlated with time.  The highest speeds are seen at the start of the flare.  Indeed, this is generally confirmable by visual inspection.  The 2002 April 21 TRACE event is a prime example of this trend.  This is not unexpected considering that the Alfv\'{e}n speed is directly proportional to the magnetic field strength which decreases with height according to the PFSS model (\citeauthor{schatten-wilcox-ness_1969}~\citeyear{schatten-wilcox-ness_1969}; \citeauthor{schrijver-derosa_2003}~\citeyear{schrijver-derosa_2003}) and the height of the reconnection site is expected to increase with time \citep{forbes-acton_1996}.  (It is also true that the Alfv\'{e}n speed is inversely proportional to the square root of the density which decreases with height; however, the progression of the flare may lead to an enhanced density in the current sheet through processes such as chromospheric evaporation and conduction \citep{reeves_2010}.  Consequently, the decrease in the field strength likely dominates the Alfv\'{e}n speed profile.)  So while the flow speeds are not near the expected Alfv\'{e}n speed, their decrease with height follows the expected Alfv\'{e}n speed trend, and it should be restated that precise measurements of the coronal magnetic field are unavailable at this time.  Changes in loop geometry as it relaxes, which affects the tension force acting on the retracting loops, has not been taken into account with this simplistic check on the speeds. 

Figure~\ref{trend_limb_selected} (e) displays a strong positive correlation, present for all instruments, for flow observations at higher initial heights as the flare progresses, which can be satisfied by at least two explanations:  

1)  As the flare progresses, the reconnection site travels upwards as field lines reconnect higher along the current sheet.  Using this explanation, however, would imply that the flows are being observed forming at the reconnection site.

2)  As the flare progresses, more hot plasma is being conducted into the current sheet thereby brightening it at increasing heights \citep{reeves_2010}.  The hot plasma then provides a bright background at increasing heights against which to observe the faint downflows. 

While option 1 would be a nice verification of the standard CSHKP model, it is unrealistic except for the aforementioned ``Cartwheel CME" flare wherein reconnection was observed to occur.  However, option 2 could provide some insight into the rate of heating within the current sheet -- a topic beyond the scope of the current paper.

Finally, Figure~\ref{trend_limb_selected} (f) shows the lack of a correlation between acceleration and initial area.  This result could affect potential drag models which is introduced in Section~\ref{sadsiisec:drag}.  Only flows that were measured using the automatic threshold technique and tracked in at least 5 frames are displayed.

\clearpage

\subsection{\label{sadsiisec:magnetic}Magnetic Derivations}

\begin{figure}[!ht] 
\begin{center}

\includegraphics[width=0.49\textwidth]{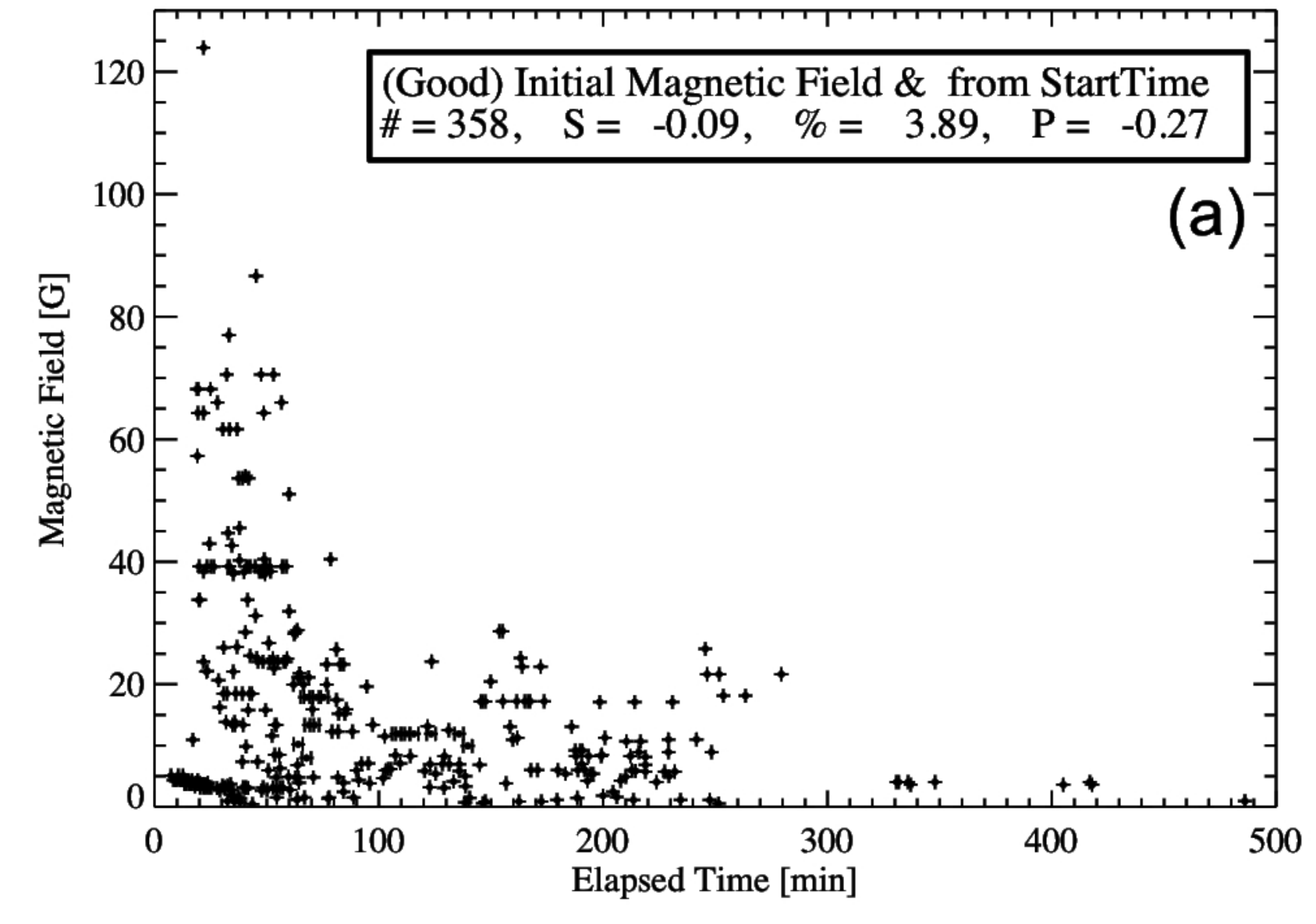}
\includegraphics[width=0.49\textwidth]{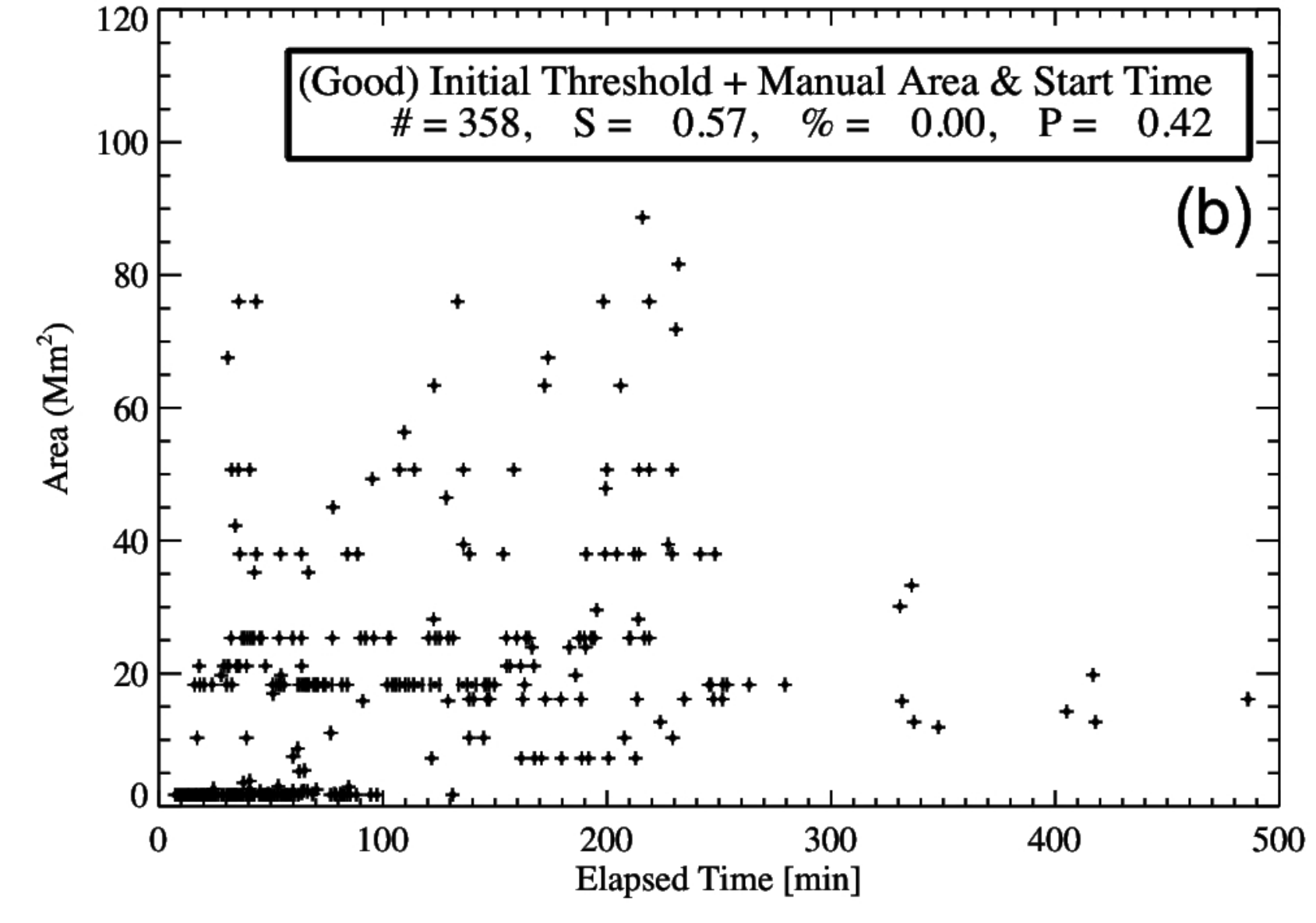}
\includegraphics[width=0.49\textwidth]{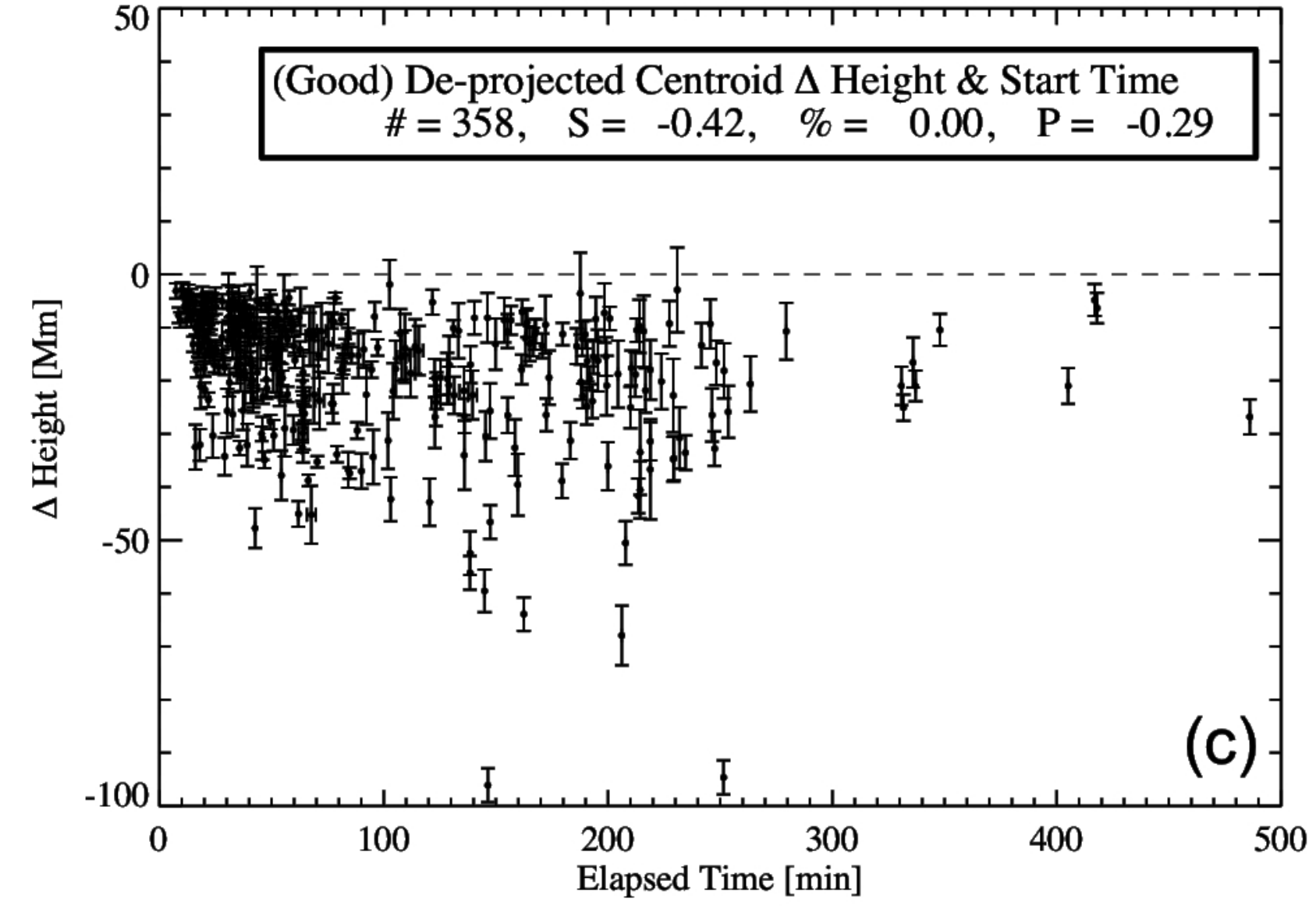}
\includegraphics[width=0.49\textwidth]{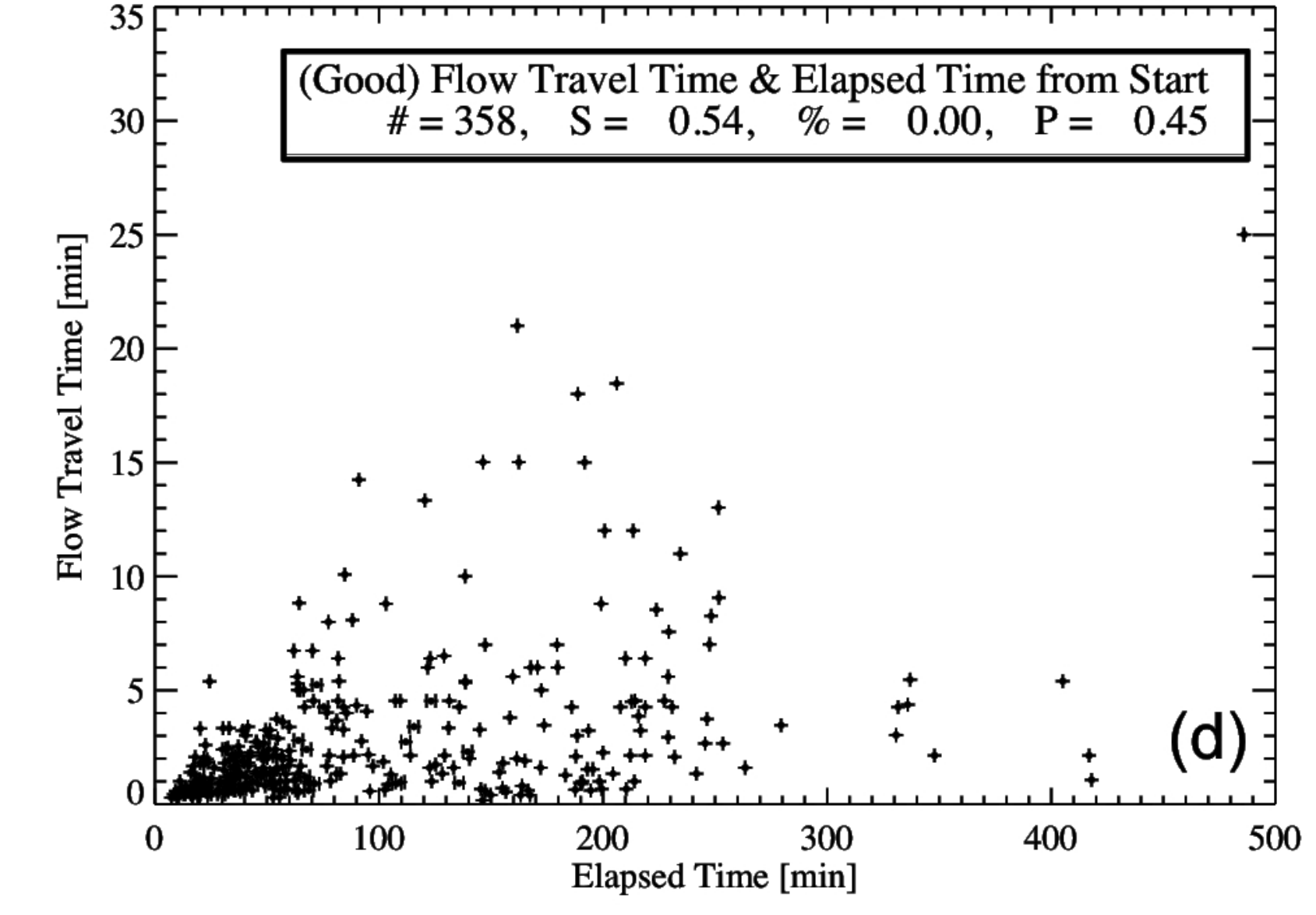}
\includegraphics[width=0.49\textwidth]{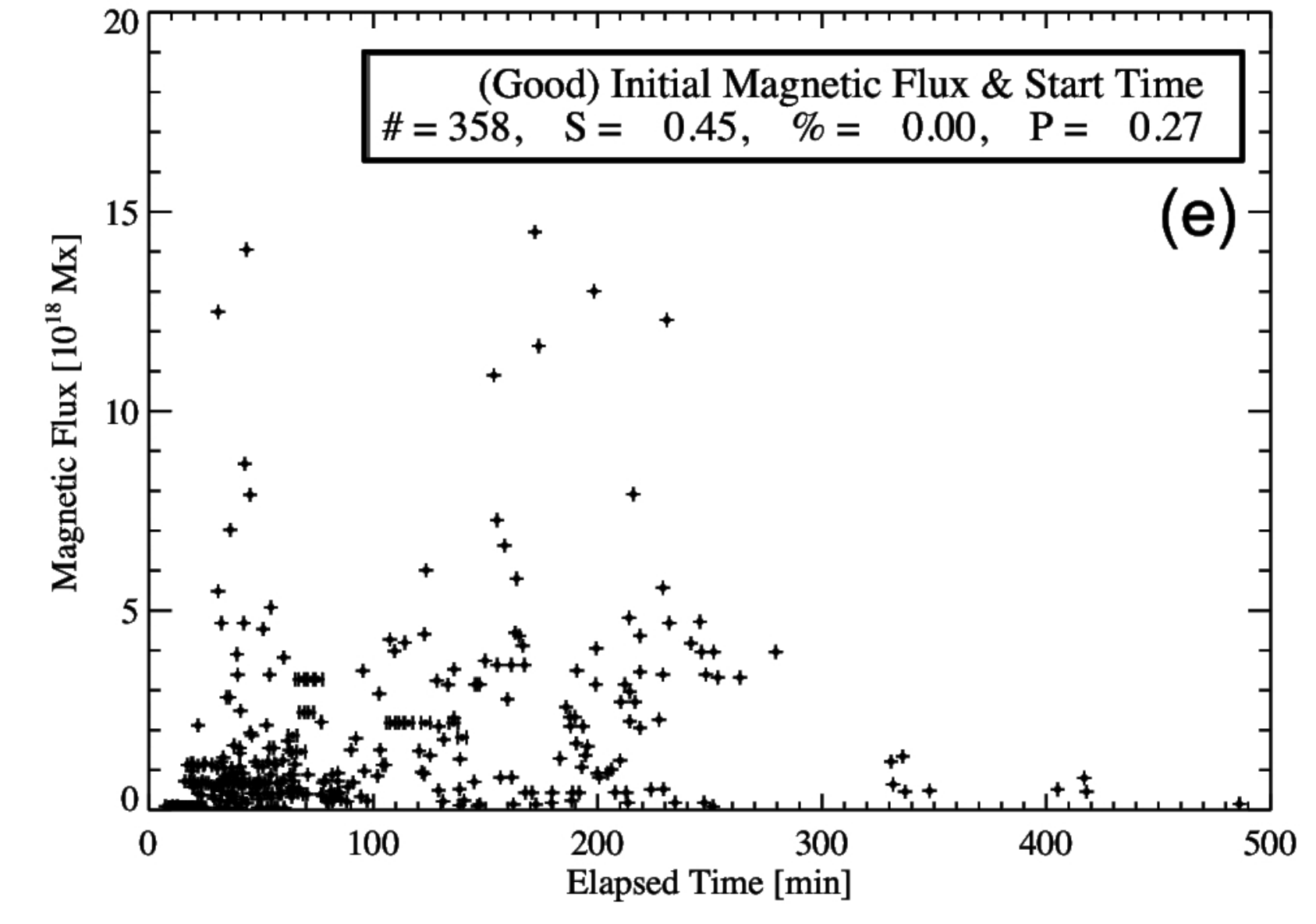}
\includegraphics[width=0.49\textwidth]{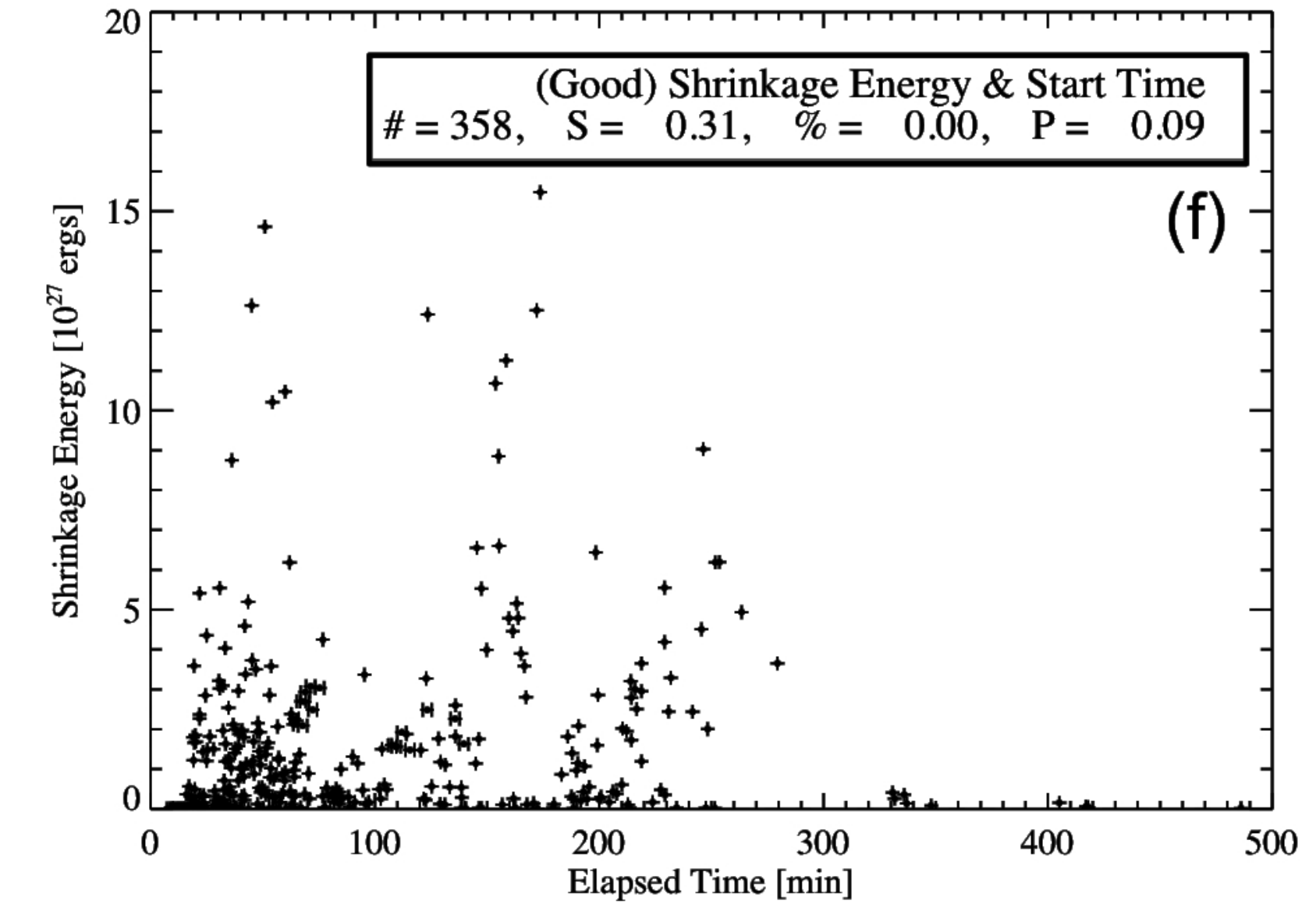}

\caption{Trends with respect to elapsed time related to the calculation of shrinkage energy.  Specifically, $\Delta$W~$\rightarrow$~(f)~$\propto$~((a)$^{2}$~$\times$~(b)~$\times$~(c))~/~(d) and $\Phi$~$\rightarrow$~(e)~$=$~(a)~$\times$~(b).  (a)  Initial magnetic field strength.  (b)  Initial area for troughs measured using the threshold technique and those measured manually.  (c)  Change in de-projected height.  (d) Flow travel time.  (e)  Initial magnetic flux.  (f)  Shrinkage energy.  The elapsed time is the time from the flare start time as indicated by the GOES light curves.  Contained with the legends are the number of flows used to create the plot (\#), the Spearman rank order correlation coefficient (S), the Spearman chance probability (\%), and the Pearson linear correlation coefficient (P).}

\label{trend_limb_work}
\end{center}
\end{figure}

Figure~\ref{trend_limb_work} provides a graphical display of the derivation of the magnetic flux and shrinkage energy with respect to elapsed time.  The elapsed time is the time (in minutes) from the flare start time as indicated by the GOES light curves.  LASCO measurements are not included (see Section~\ref{sadsiisec:quartiles} for an explanation).  

While the correlation coefficients for Figure~\ref{trend_limb_work} (a) are marginal, the visual trend suggests an anti-correlation between magnetic field strength and time.  The field strength is derived using the initial height position and is therefore very sensitive to the initial detection.  Because the early flows are often detected closer to the solar surface as the supra-arcade region is brightening, their associated field strengths are strong.  Conversely, the later flows can sometimes be detected at higher initial heights (refer to Figure~\ref{trend_limb_selected} (e)) where the field strengths are lower.

Panel (b) does not support a correlation between the initial areas and elapsed flare time.  The clusters of area measurements are a result of manual measurements.  The discrete nature of the areas is due to pixelation.

The change in height is shown in panel (c).  These values are negative because the final height is lower than the initial, so while the coefficients indicate a negative correlation, the actual interpretation is that the magnitude of the change in height increases with elapsed flare time.  This result is again due to the ability to measure initial positions higher above the arcade as the flare progresses and the supra-arcade region becomes more illuminated.  The amount of time that an individual flow is tracked (panel (d)) shows a positive correlation for this same reason.

Finally, magnetic flux ($\Phi~=~B~\times~A$ or (e)~$=$~(a)~$\times$~(b)) is shown in Figure~\ref{trend_limb_work}~(e), and magnetic shrinkage energy ($\Delta~W~=~B^2~A~\Delta~L~/~8\pi$ or (f)~$\propto$~((a)$^{2}$~$\times$~(b)~$\times$~(c))~/~(d)) is shown in (f).  Neither parameter shows a strong correlation with elapsed time.  GOES SXR flare lightcurves indicate that most of the flare energy is released early followed by a slow decline in emission as the flare progresses (for long duration events); therefore, the expectation is for the shrinkage energy estimates and/or the number of flows to be anti-correlated with elapsed time from the start of the flare.  The former is not obviously supported by Figure~\ref{trend_limb_work}~(f), but this could be due to the combination of several sources of uncertainty which are difficult to include in the error bars (e.g. de-projection assumptions, footpoint assignments, magnetic model outputs, etc.).  (Plotting the shrinkage energy versus time from peak flare time shows a similar trend to that of Figure~\ref{trend_limb_work}~(f).)  The number of flows does appear to decrease during the decay phase of most flares as expected.  Precise examination of flow timings with respect to SXR and hard X-ray (HXR) flare output is currently underway.

The precise geometry of the loops is unknown and several assumptions would be required in order to more accurately estimate their change in volume.  The equation for the shrinkage energy given above is derived from $\Delta~W~=~B^2~\Delta~V~/~8\pi$, assuming a constant cross-section.  The $\Delta~L$ measurement is approximated by the path length of the flows.  Twice this value may possibly reflect a more accurate portrayal of the actual change in volume, but that assumption is not included in our calculations for Figure~\ref{trend_limb_work}~(f).  Because of the uncertainties involved with all of the flow measurements, order of magnitude estimates are more reasonable considerations.

Reconnection rates can be inferred by summing the flow fluxes and dividing by the total time of flow observations per flare.  Five flares in this study have enough reliable flow detections to make this derivation meaningful.  Reconnection rates reported in \cite{mckenzie-savage_2009} have been updated (due to processing of more flows with better precision) for the 1999 January 20, 2000 July 12, and 2002 April 21 flares to $0.8~\times~10^{16}$, $2.3~\times~10^{16}$, and $1.2~\times~10^{16}$~Mx s$^{-1}$, respectively.  For the 2002 July 23 TRACE flare, $\sim0.3~\times~10^{19}$ Mx was processed in 55~minutes, for a reconnection rate of $0.1~\times~10^{16}$~Mx~s$^{-1}$.  Similarly, for the 2003 November 4 TRACE flare, $\sim4.1~\times~10^{19}$ Mx was processed in 41~minutes, for a reconnection rate of $1.7~\times~10^{16}$~Mx~s$^{-1}$.  The rates obtained for the 1999 January 20 and 2002 July 23 flares are likely underestimates because they occurred on the east limb.  The east limb magnetic field measurements have been established as being consistently low due to the lack of current magnetograms (see Figures~\ref{radial_field} \&~\ref{quart3} (a)).  Reconnection rates are not being reported for the 2008 April 9 flare despite its significant number of detectable flows because it occurred well beyond the limb making the magnetic field estimates unreliable.  These reported reconnection rates may represent a lower bound considering that we can only include the flux and energy from detectable and trackable flows.  Additional flows within the same time frame would increase the energy budget and, in turn, the reconnection rate.

\clearpage

\subsection{\label{sadsiisec:drag}Drag Analysis}

Based on often-assumed estimates for coronal Alfv\'{e}n speeds of about 1000 km~s$^{-1}$, the flows seem to be moving well below predictions.  This contradiction may not actually be the case considering that coronal Alfv\'{e}n speeds cannot be well constrained.  The signal is simply too low to measure the magnetic field or density precisely in the supra-arcade region.  However, if the Alfv\'{e}n speed is indeed on order of 1000 km~s$^{-1}$, then the slow flow speeds can be attributed to at least two possibilities:  1) initial height and 2) drag.  

The flows are expected to be accelerated from the reconnection site (the duration of acceleration is model-dependent) and slow as they settle near the top of the post-eruption arcade.  The fact that the initial height estimates are associated with large uncertainty due to the lack of long exposure duration images of the supra-arcade region has been discussed in previous sections.  Therefore, if the flows are initially observed far below the reconnection site, then the measured speed is likely to be slower than the actual initial speed.  Note, however, that the measured speeds are still relatively slow (\textless~200~km~s$^{-1}$) for the ``Cartwheel CME" event near the reconnection region \citep{savage_2010}.  This bias has been well established, so in this section we will explore the second option -- drag.

\begin{figure}[!ht] 
\begin{center}

\includegraphics[width=0.49\textwidth]{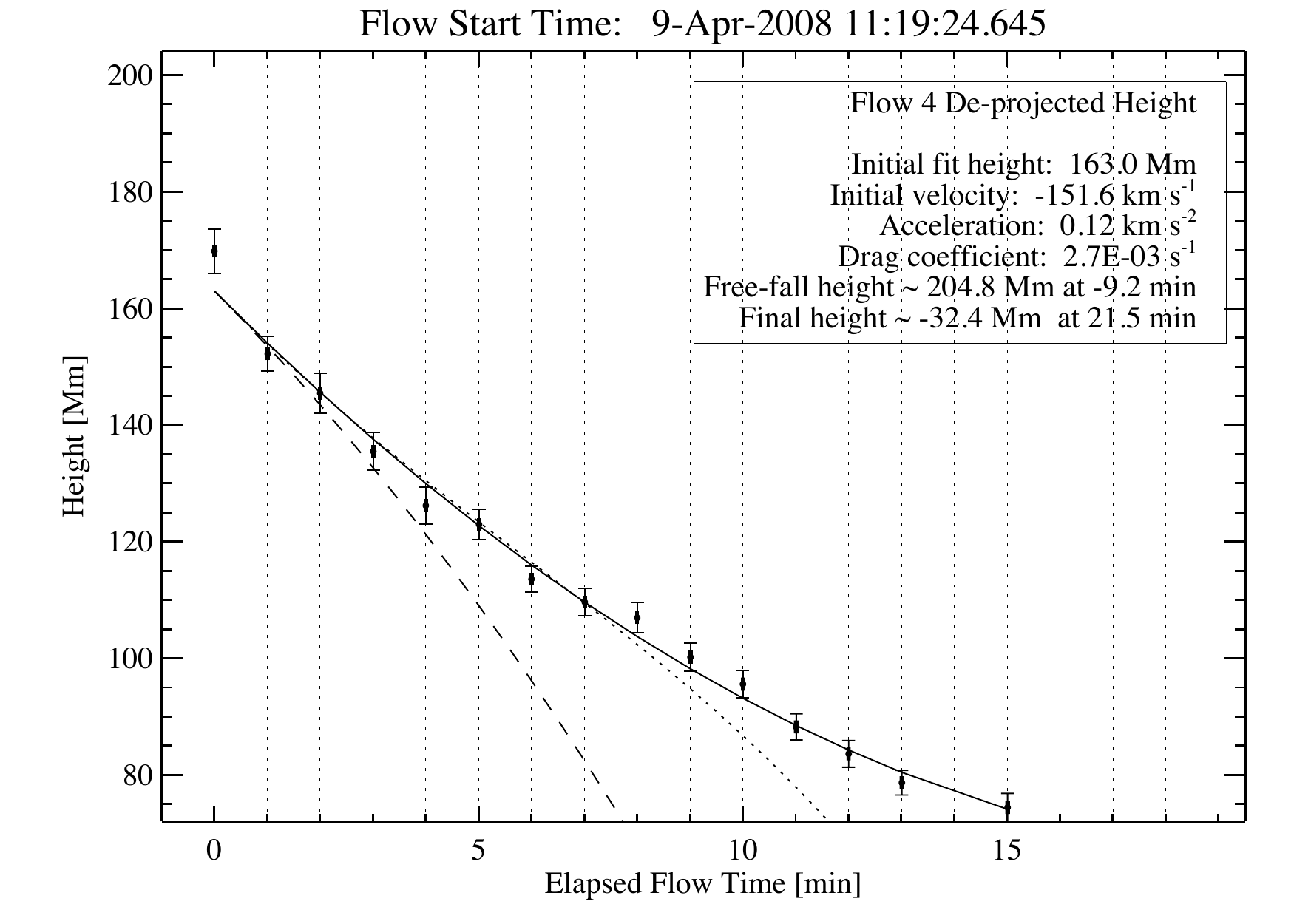}
\includegraphics[width=0.49\textwidth]{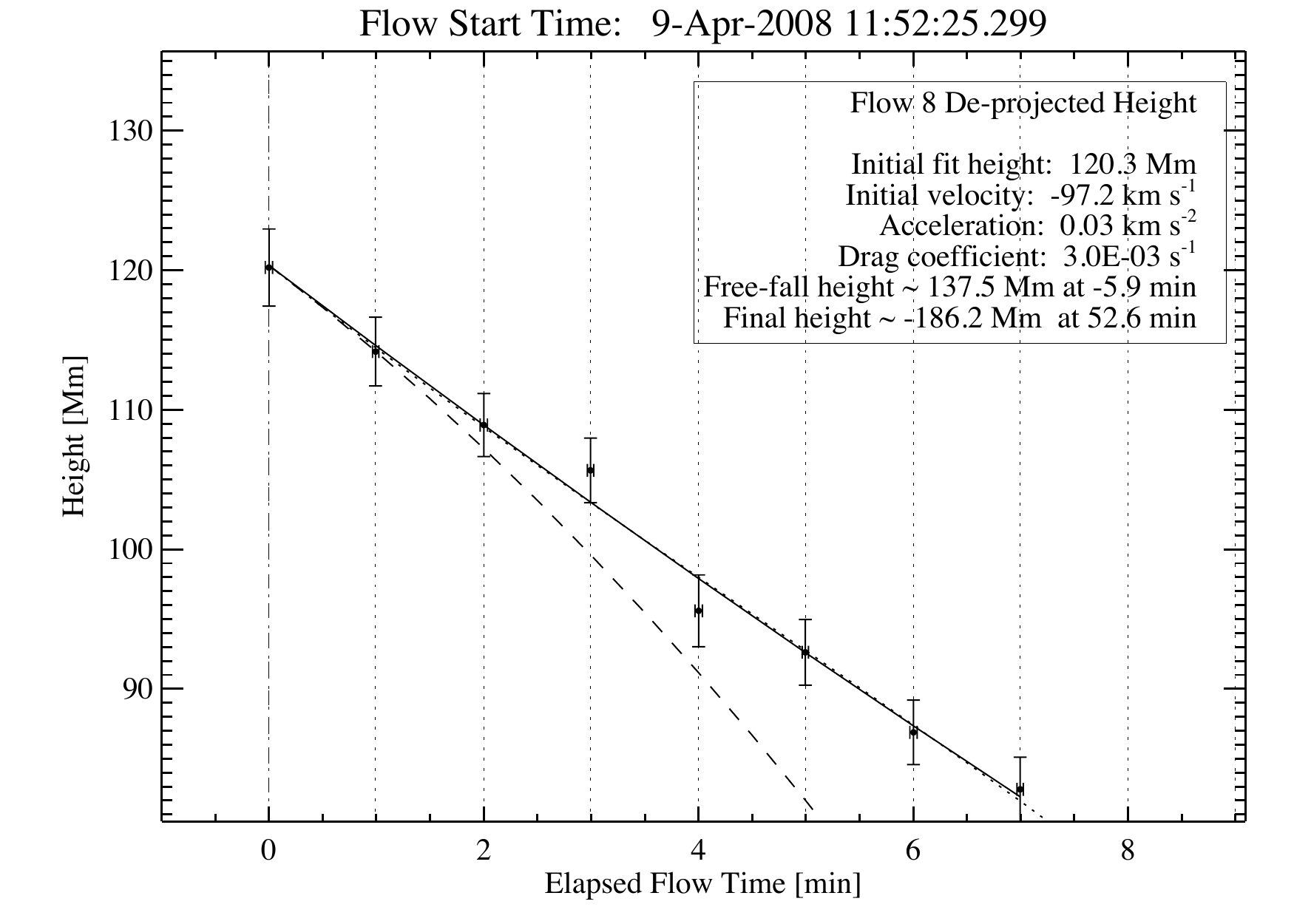}

\caption{Example flow profiles from the 2008 April 9 ``Cartwheel CME" event observed by XRT.  The thick line represents the 2-D polynomial fit to the trajectory.  The thick dashed line represents the gravitational profile calculated using the initial height and initial speed of the flow.  The thin dashed line represents the gravitational profile with some constant damping coefficient applied.  The thin dashed vertical lines indicate the times of the available images.}

\label{gravity_profiles}
\end{center}
\end{figure}

One way to test the effects of drag on the flows is to compare the trajectories to theoretical gravitational profiles.  While the flows are not expected to be affected by gravity if they are plasma-deficient flux tubes, damped gravitational profiles provide a method of comparing a known trajectory affected by drag with the shape of the flow profiles.  This test was done for the flows from the ``Cartwheel CME" event. Figure~\ref{gravity_profiles} provides a few example flow profiles with overlaid gravitational profiles as a thick dashed line.  

The gravitational profiles were calculated using the flow initial height (h$_{0}$) and speed (v$_{0}$) (explicitly, $h = h_{0} + v_{0} t + 0.5 g_{\odot} t^{2}$, where t = time and g$_{\odot}$ = acceleration due to gravity calculated as $g_{\odot} = -(6.67\times10^{-11}~M_{\odot}) / ((R_{\odot} + h_{0})^{2}$).  The thin dashed line is the gravitational profile with some constant damping in the form of Stokes' drag ($F_{d} = -b v$) where the speed is thus calculated as $v = g_{\odot}/k + ((k v_{0} - g_{\odot})/k) exp(-k t)$, where k = damping coefficient.  This form of drag is considered based on its simplicity for this exercise.  The drag coefficient is noted in the legend and was chosen as the value resulting in a profile closest to that of the original flow trajectory.  A few of the flow profiles from this event match well with the gravitational profiles (e.g. Figure~\ref{gravity_profiles} (c)), but most begin to veer away towards the end of the track.  Those tracks that do not appear to decelerate may be too incomplete to derive a proper acceleration.  Additionally, the entire arcade is obscured by the limb for this event which makes following the flows until they reach their final configuration impossible.

The initial height from free-fall (drag-free) is noted in the legend along with a predicted final height for decelerating flows.  No convergence among these values was found which is most likely due to the inability to observe and measure complete flow tracks thus resulting in imprecise profiles.  Also, the applied 2-D polynomial fit does not take into account changes in the acceleration which has a noticeable effect.  

\begin{figure}[!ht] 
\begin{center}

\includegraphics[width=0.6\textwidth]{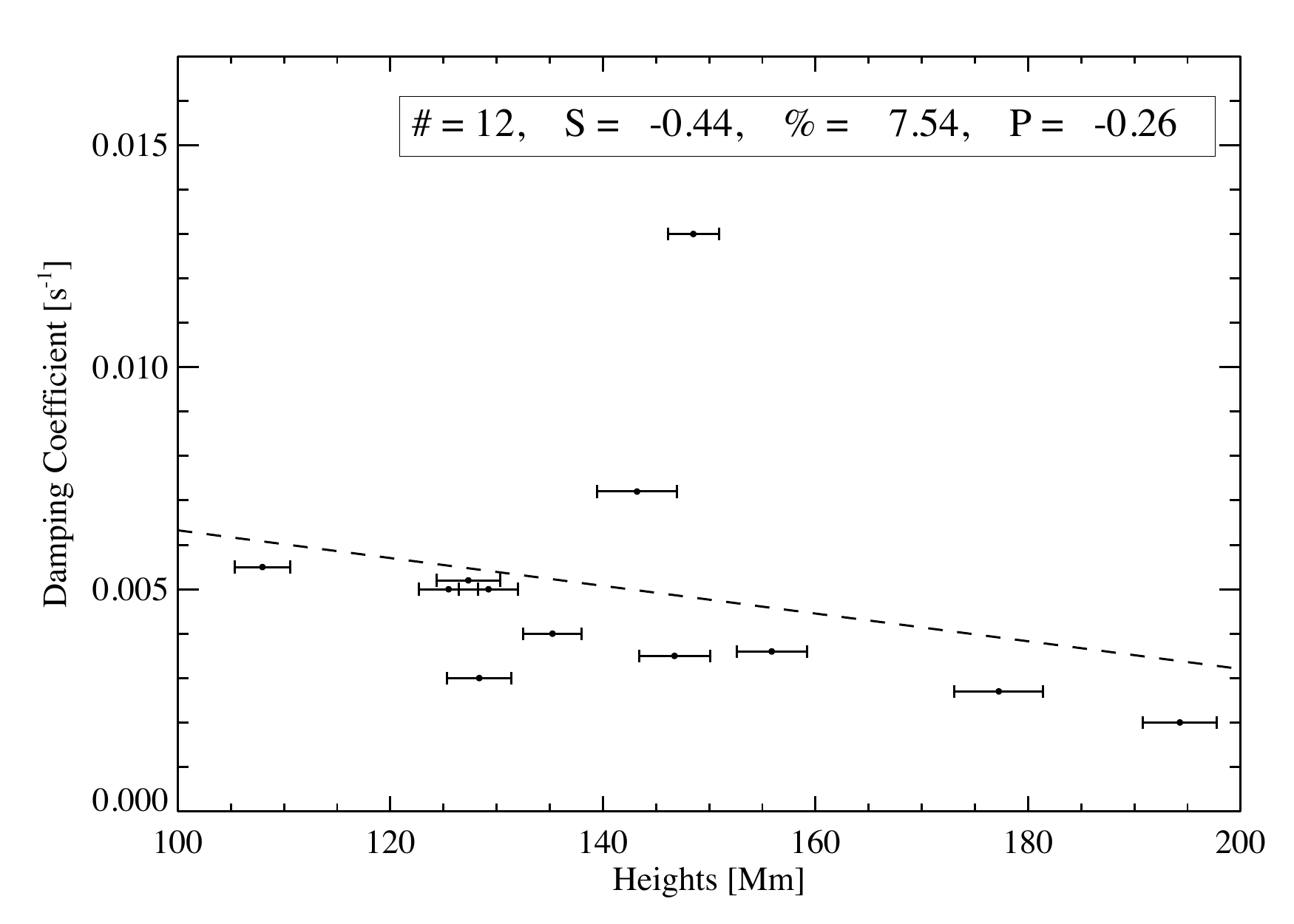}

\caption{A plot of the applied drag constants, chosen to most closely match the flow profiles, versus initial height.  The dashed line represents a linear fit to the data.  Contained with the legend is the number of flows used to create the plot (\#), the Spearman rank order correlation coefficient (S), the Spearman chance probability (\%), and the Pearson linear correlation coefficient (P).}

\label{drag_height}
\end{center}
\end{figure}

Considering how well the damped profiles match with the flow profiles, drag is a possible explanation to the slow-speed problem.  As noted above, constant drag is used to calculate the thin dashed profiles, but realistically the drag should be a function of height if it is due to the flows entering regions of higher density.  Figure~\ref{drag_height} supports the anti-correlation between height and drag with a Spearman rank order correlation coefficient of -0.44 and a linear correlation coefficient of -0.26.  Interestingly though, Figure~\ref{trend_limb_selected}~(f) does not indicate any correlation between acceleration and area.  If the drag is a result of mass build-up in front of the flow, the expectation would be for the flows with larger areas to slow faster.  This could possibly be an indicator of the imprecision associated with flow area measurements.  If so, then drag models using mass build-up need to be assessed using parameters independent of area.  

Some interpretations of these results are that either the Alfv\'{e}n speed is lower than expected in the reconnection region, the drag may be overwhelming the magnetic tension force pulling down the loops causing them to retract slower, entanglement of field lines or plasma compressibility during the reconnection process reduces the initial speed, or even continual field entanglement during the retraction phase through the current sheet slows the flows.  There is also the possibility that the flows are not retracting reconnected loops; however, several of the flare observations clearly show shrinking loop features, and the slowing of the flows as they approach the arcade would become difficult to explain otherwise.

\clearpage

\subsection{\label{sadsiisec:loop_cooling}Loop Cooling}

\begin{figure}[!ht] 
\begin{center}

\includegraphics[width=0.8\textwidth]{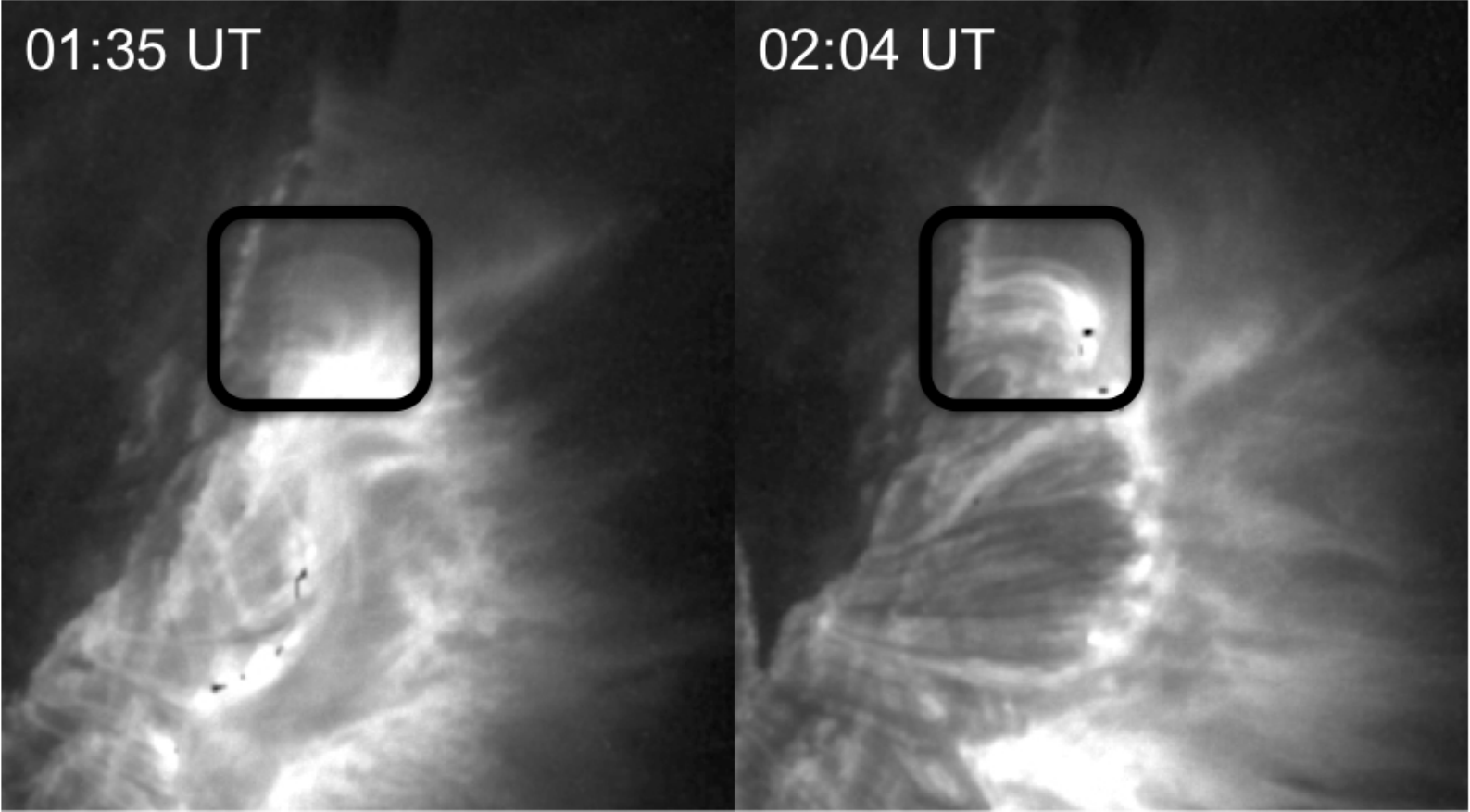}

\caption{Images taken from the 2002 April 21 TRACE flare showing some of the post-eruption arcade loops cooling between the 195 \AA\ temperature passbands.}

\label{loop_cooling}
\end{center}
\end{figure}

An interesting observation can be seen to occur during the 2002 April 21 and 2003 November 4 TRACE events wherein faint, hazy loops retract early during the impulsive phase along the length of the arcade.  Between approximately 15$-$40 minutes later, bright, distinguishable post-eruption arcade loops appear without retracting in the same region.  Figure~\ref{loop_cooling} provides a visual reference of these loops and the delay.  Deriving a more precise figure for this delay is complicated by the fact that a one-to-one correspondence cannot be established between a shrinking loop and an arcade loop.  This delay is not observable in XRT images.  Its appearance using TRACE is probably due to the use of the 195 \AA\ filter which selects for temperatures of 0.5$-$2 MK and 11$-$26 MK.   Therefore, while the early hot retracted loops have already settled, it appears to take the plasma within them, which is assumed to be supplied through chromospheric evaporation, approximately 15$-$40 minutes to cool into the lower temperature passband.  This has interesting implications:  namely, that the early shrinking loops are either very hot to begin with, or that they quickly fill with hot plasma as they reach their potential configuration above the arcade.  This delay cannot possibly be measured in the SXT data because the arcade region is saturated.

\clearpage

\subsection{\label{sadsiisec:discussion}Discussion \& Conclusions}

The preceding sections have presented many downflow observations from several instruments.  Measurements of any one flow have too many possible sources of uncertainty to contribute to our understanding of the reconnection process occurring during long duration solar flares.  SXT observed the most flares during its lifetime because it was operational throughout two solar maxima and was the first instrument to observe SADs; however, the poor spatial resolution leaves much to be desired with respect to being able to observe many flows per flare and make reliable measurements.  LASCO has observed a number of flows in the outer corona; however, they are not always associated with flaring events, and because of the imager's much higher observational regime above the solar surface and considerably lower resolution, comparisons with other SADs and SADLs observations are difficult to interpret.  LASCO speeds and accelerations are the only parameters comparable to the other instruments.  TRACE has the best resolution of all the instruments used for this study and has observed many hundreds of flares during its lifetime; however, flows are difficult to observe with TRACE unless the 195 \AA\ filter is being utilized for broader temperature coverage and the flare is atypically large (all three TRACE flares in this study were GOES X-class flares).  These requirements are due to the need for the hot plasma above the flare arcade to be illuminated.  Finally, XRT has high enough spatial resolution to observe the flows as well as the optimal temperature coverage; however, solar activity has been unusually low throughout most of its lifetime to date making the amount of available flare data small.  For all of these reasons, combining flow observations from all instruments improves our understanding of the flows themselves and ultimately our understanding of the reconnection process.  Also, comparing the measurements between the instruments allows us to determine if the appearance of flows is temperature or density dependent.

Interpreting SADs as the cross-sections of retracting reconnected flux tubes also means that if they are viewed from an angle that is not near perpendicular to the arcade axis (i.e. the polarity inversion line), the downflows will instead appear as shrinking loops.  These shrinking loops (SADLs) have indeed been clearly observed with all of the instruments under investigation.  Therefore, comparing observations of SADs to those of SADLs can help to support or refute the hypothesis for SADs.

Figures~\ref{quart1} through \ref{quart3} present a summary of the instrument and SAD/SADL comparisons.  These figures show that the flow velocities and accelerations agree between the instruments quite well.  Height measurements agree except for those measured with XRT due to the exceptional heights observed for the ``Cartwheel CME" flare.  Figure~\ref{quart2} (a) shows that the area measurements are understandably resolution dependent, which indicates that we may not be able to observe the smallest loop sizes.  The flux and energy measurements are area dependent and therefore instrument dependent.  There is also a limb dependence with the magnetic measurements due to the use of modeling based on magnetograms.  Even so, there is decent agreement between all of the instruments.  (LASCO is only included with the velocity and acceleration comparisons as explained in Section~\ref{sadsiisec:quartiles}.)

\begin{figure}[!ht] 
\begin{center}
\includegraphics[width=.7\textwidth]{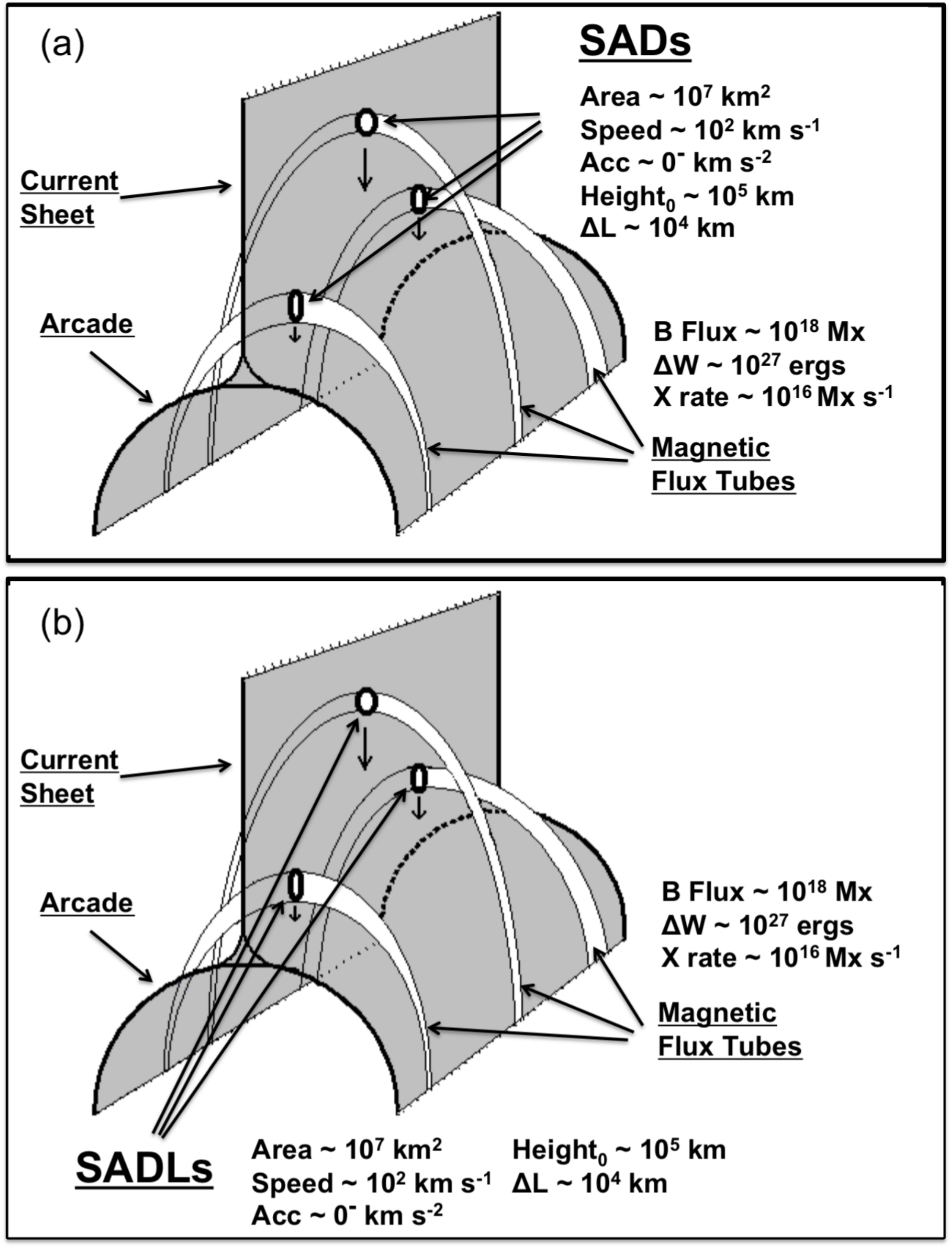}
\caption{(a)  Schematic diagram of supra-arcade downflows (SADs) resulting from 3-D patchy reconnection.  Discrete flux tubes are created, which then individually shrink, dipolarizing to form the post-eruption arcade.  The measured quantities shown are averages from the limb events listed in Table~\ref{sads_list_analyzed} containing SADs.  (b)  Schematic diagram of supra-arcade downflowing loops (SADLs) also resulting from 3-D patchy reconnection.  The measured quantities shown are averages from the limb events listed in Table~\ref{sads_list_analyzed} containing SADLs.}
\label{sads_sadls_diagram_mag}
\end{center}
\end{figure}

Beyond the agreement between the SADs and SADLs measurements, the high-resolution TRACE observations clearly show both SADs and SADLs occurring during the same flare depending on the arcade viewing angle which curves within the active region.  The SADs versus SADLs diagram shown in Figure~\ref{sads_sadls_diagram_mag} (previously provided in \cite{mckenzie-savage_2009} -- Figure~10 \& \cite{savage_2010} -- Figure~23) still remains applicable after this analysis and has been updated to include SADLs with magnetic estimates.  

The measured cross-sectional areas range from $\sim$2--90~Mm$^{2}$, with at least 75\% being smaller than 40~Mm$^{2}$ (Figure~\ref{quart2}~(a)).  The flows typically move at speeds on order of 10$^{2}$~km~s$^{-1}$ with accelerations that are near zero or slightly decelerating.  The most complete flow paths show significant deceleration near the top of the arcade.  There is a range of initial heights depending on the quality of the image set, but they are generally about 10$^{5}$~km above the solar surface with a path length of $\sim$10$^{4}$~km.  Each tube carries $\sim$10$^{18}$~Mx of flux and releases on order of 10$^{27}$~ergs of energy as it retracts.  A lower limit of 10$^{16}$~Mx~s$^{-1}$ can be put on the reconnection rate by considering  the total flux released by the observed flows for 5 flares (Section~\ref{sadsiisec:magnetic}).

These observations and measurements support the conclusion that SADs are indeed post-reconnection loops relaxing to form the post-eruption arcade.  Also, the lack of instrument dependency of the dark flow observations suggests that either the loops are filled with cold material or are depleted.  The temperature coverage of the instruments used in this study goes up to about 100 MK; therefore, it is unlikely that the loops are filled with hotter material.  Combining this with the SUMER analysis of the 2002 April 21 TRACE flare from \cite{innes-mckenzie-wang_2003a} which showed the lack of continuum absorption or emission in the C II, Fe XII, and Fe XXI lines at the flow sites supports the hypothesis that the tubes are depleted.  This is at least true for loops reconnecting following the flare impulsive phase, which may be due to the fact that, according to the standard model, subsequent loops reconnect higher in the corona where less plasma is available to fill the loops.  Loops shrinking very early during the 2002 April 21 TRACE flare appear to be bright in the hot 195 \AA\ bandpass as noted in Section~\ref{sadsiisec:loop_cooling}.  Few bright flows are analyzed in this study because they are more difficult to observe as SADs due to the low contrast although bright SADLs have easily been observed (see Figure~\ref{cartoon_bright_dark}).  Density analysis has not been performed due to the small sample of bright, tracked flows and especially lack of spectral coverage.

\clearpage

There are a few trends to note from Figure~\ref{trend_limb_selected}.  Panel (a) shows that SADs tend to have increased areas in regions of weaker magnetic field.  This result may be due to the fact that the weaker magnetic fields are generally associated with higher coronal heights where the signal to noise is very low.  The noise impedes precise area measurements.  If the correlation is real though, it may indicate that a given reconnection episode is associated with a limited amount of flux transfer.  There is also the possibility that the loop cross-sections shrink as they retract either physically or apparently (due to filling with heated plasma from chromospheric evaporation).  In this situation, lower initial heights, where the magnetic field strength is larger, would be associated with smaller areas as well.  Then in panel (e), the initial heights seem to increase with time which is an expected consequence from the CHSKP model.  This trend, however, is likely the result of an observational bias due to background brightening as the flare progresses rather than evidence of an upwardly migrating X-point.

The observational findings presented in this paper provide a more complete description of the SAD/SADL phenomenon than has previously been available.  Assuming that SADs and SADLs are thin, post-reconnection loops based on this body of evidence, the measurements obtained through this analysis and summarized in Figure~\ref{sads_sadls_diagram_mag} provide useful constraints for reconnection models.  Area estimates can constrain the diffusion time per episode and reconnection rates can be derived to distinguish between fast and slow reconnection.  Creation of outflowing flux tubes carrying on order of 10$^{18}$~Mx of flux, with net reconnection rates of at least 10$^{16}$~Mx~s$^{-1}$, should be an objective of realistic models of 3D reconnection.  The lack of acceleration of the downflow speeds and their discrete nature tends to favor 3D patchy Petschek reconnection.  Speeds almost an order of magnitude slower than traditionally assumed Alfv\'{e}n speeds are an unexpected consequence of the flow measurements; therefore, analyzing the effect of some source of drag on the downflow trajectories using models (an effort begun by \cite{linton-longcope_2006}) could provide valuable insight into this discrepancy.  

\subsection*{Acknowledgements}

This work was supported by NASA under contract NNM07AB07C with the Harvard-Smithsonian Astrophysical Observatory.  The authors would like to thank Drs. D. Longcope, C. Kankelborg, J. Qiu, and A. Des Jardins for constructive conversations.  Hinode is a Japanese mission developed and launched by ISAS/JAXA, with NAOJ as domestic partner and NASA and STFC (UK) as international partners.  It is operated by these agencies in cooperation with ESA and NSC (Norway).  Yohkoh data are provided courtesy of the NASA-supported Yohkoh Legacy Archive at Montana State University.

\begin{center}
\begin{sidewaystable}
\caption{List of flares exhibiting downflow signatures.}
\label{sads_list_full}
\footnotesize
\begin{tabular*}{\textwidth}{@{\extracolsep{\fill}}  l  c  c  c  c  c  c l  c}
\\
\toprule
{\textbf{\#}} & \textbf{YYYYMMDD} & {\textbf{Approx. Time}} & {\textbf{AR}}  & {\textbf{GOES}} & {\textbf{FOV Coords}} & {\textbf{Instrument}}  & {\textbf{Filter}} & {\textbf{Analyzed}} \\
\hline
1 & 19911216  & 12:30 - 14:30 & 06972 & M3.2 & E88 S08 & SXT & H-Al12 &  \\
2 & 19920731  & 00:45 - 05:30 & 07244 &           & E05 S00 & SXT & Q-Al.1 &  \\
3 & 19921102  & 05:00 - 11:00 & 07321 & X1.9 & W90 S27 & SXT & H-Al.1 &  \\
4 & 19930514  & 22:00 - 00:00 & 07500 & M4.4 & W50 N17 & SXT & H-AlMg &  \\
5 & 19930624  & 07:45 - 08:30 & 07529 & M9.7 & E74 S14 & SXT & H-AlMg &  \\
6 & 19940227  & 09:00 - 10:00 & 07671 & M2.8 & W88 N09 & SXT & H-AlMg &  \\
7 & 19980420  & 09:15 - 11:00 & 08202 & M1.4 & W92 S23 & SXT & H-AlMg &  \\
8 & 19980423  & 05:30 - 07:15 & 08210 & X1.2 & E91 S18 & SXT & H-AlMg & \checkmark \\
9 & 19980427  & 09:30 - 12:00 & 08210 & X1.0 & E54 S17 & SXT & H-AlMg & \checkmark \\
10 & 19980506  & 07:30 - 10:00 & 08210 & X2.7 & W65 S12 & SXT & H-AlMg & \checkmark \\
11 & 19980509  & 03:00 - 06:00 & 08210 & M7.7 & W91 S19 & SXT & H-AlMg & \checkmark \\
12 & 19980816  & 18:30 - 19:30 & 08306 & M3.2 & E94 N30 & SXT & H-AlMg & \checkmark \\
13 & 19980818  & 22:15 - 00:30 & 08306 & M5.4 & E94 N32 & SXT & H-AlMg &  \\
14 & 19980920  & 02:30 - 03:30 & 08340 & M1.9 & E62 N20 & SXT & H-AlMg &  \\
15 & 19980930  & 13:15 - 14:15 & 08340 & M2.9 & W92 N20 & SXT & H-AlMg & \checkmark \\
16 & 19981123  & 11:00 - 12:30 & 08392 & M3.2 & E60 S24 & SXT & H-AlMg & \checkmark \\
17 & 19981223  & 05:45 - 07:45 & 08421 & M2.3 & E89 N23 & SXT & H-AlMg & \checkmark \\
18 & 19990120  & 19:00 - 23:00 & 08446 & M5.2 & E87 N29 & SXT & H-AlMg & \checkmark \\
19 & 19990216  & 13:15 - 04:15 & 08458 & M3.3 & W17 S27 & SXT & H-AlMg &  \\
20 & 19990503  & 06:00 - 07:00 & 08530 & M4.4 & E47 N20 & SXT & H-AlMg & \checkmark \\
21 & 19990508  & 11:15 - 11:30 & 08541 & M1.6 & E78 N20 & SXT & H-AlMg &  \\
22 & 19990508  & 14:30 - 15:00 & 08526 & M4.7 & W80 N19 & SXT & H-AlMg &  \\
23 & 19990511  & 21:45 - 22:15 & 08542 & C1.6 & E87 S20 & SXT & H-AlMg &  \\
24 & 19990725  & 13:00 - 14:00 & 08639 & M2.4 & W93 N36 & SXT & H-AlMg & \checkmark \\
25 & 19991128  & 18:15 - 22:15 & 08771 & M1.6 & W89 S14 & SXT & H-AlMg & \checkmark \\
26 & 19991207  & 01:00 - 01:30 & 08781 & C8.7 & W89 S10 & SXT & H-AlMg &  \\
27 & 20000101  & 15:00 - 00:00 &  &  & E N & LASCO & O+C & \checkmark \\
28 & 20000222  & 20:15 - 21:30 & 08882 & M1.1 & E82 S19 & SXT & H-AlMg & \checkmark \\
29 & 20000712  & 02:30 - 03:45 & 09087 & C5.3 & E89 S11 & SXT & H-AlMg & \checkmark \\
30 & 20000712  & 21:00 - 22:00 & 09066 & M1.9 & W86 N15 & SXT & H-AlMg & \checkmark \\
31 & 20000930  & 17:45 - 18:45 & 09178 & M1.0 & E85 S32 & SXT & H-AlMg &  \\
\hline
\end{tabular*}
\end{sidewaystable}
\end{center}

\begin{center}
\begin{sidewaystable}
Table~\ref{sads_list_full} Continued.\\ \\
\footnotesize
\begin{tabular*}{\textwidth}{@{\extracolsep{\fill}}  l  c  c  c  c  c  c l  c}
\toprule
{\textbf{\#}} & \textbf{YYYYMMDD} & {\textbf{Approx. Time}} & {\textbf{AR}}  & {\textbf{GOES}} & {\textbf{FOV Coords}} & {\textbf{Instrument}}  & {\textbf{Filter}} & {\textbf{Analyzed}} \\
\hline
32 & 20001016  & 06:30 - 07:30 & 09193 & M2.5 & N05 W75 & SXT & H-AlMg & \checkmark \\
33 & 20001025  & 14:00 - 18:00 & 09199 & C2.1 & W69 N10 & SXT & F-AlMg &  \\
34 & 20001108  & 23:15 - 00:15 & 09213 & M7.5 & W90 N10 & SXT & H-AlMg & \checkmark \\
35 & 20001125  & 01:00 - 02:30 & 09240 & M8.3 & E51 N08 & SXT & H-AlMg &  \\
36 & 20010119  & 17:15 - 18:15 & 09313 & M1.2 & E58 S08 & SXT & H-AlMg &  \\
37 & 20010402  & 21:15 - 22:00 & 09393 & X20 & W70 N16 & SXT & H-AlMg & \checkmark \\
38 & 20010402  & 23:45 - 02:30 & 09393 & M1.2 & W70 N16 & SXT & H-AlMg & \checkmark \\
39 & 20010403  & 03:30 - 07:15 & 09415 & X1.2 & E89 S22 & SXT & H-AlMg & \checkmark \\
40 & 20010404  & 10:00 - 12:00 & 09415 & M1.6 & E59 S22 & SXT & H-AlMg &  \\
41 & 20010405  & 20:45 - 23:30 & 09415 & M5.1 & E47 S21 & SXT & H-AlMg &  \\
42 & 20010626  & 15:00 - 19:00 &  &  & E90 S20 & SXT & F-AlMg & \checkmark \\
43 & 20010825  & 16:30 - 17:00 & 09591 & X5.4 & E28 S18 & SXT & H-AlMg &  \\
44 & 20010927  & 10:00 - 15:00 & 09628 & M1.0 & W39 S18 & SXT & F-AlMg & \checkmark \\
45 & 20011001  & 04:30 - 11:30 & 09632 & M9.1 & W75 S18 & SXT & H-AlMg &  \\
46 & 20011009  & 11:00 - 11:30 & 09653 & M1.4 & E11 S22 & SXT & H-AlMg & \checkmark \\
47 & 20011030  & 19:00 - 21:00 & 09687 & C5.0 & E90 S19 & SXT & H-AlMg & \checkmark \\
48 & 20011101  & 14:00 - 17:30 & 09687 & M1.8 & E90 S19 & SXT & H-AlMg & \checkmark \\
49 & 20011109  & 18:30 - 19:15 & 09687 & M1.9 & W31 S19 & SXT & H-AlMg & \checkmark \\
50 & 20011214  & 09:45 - 10:15 & 09742 & M3.6 & E90 N09 & SXT & H-AlMg &  \\
51 & 20020421  & 01:00 - 03:00 & 09906 & X1.5 & W91 S14 & TRACE & 195 & \checkmark \\
52 & 20020723  & 00:15 - 01:30 & 10039 & X4.8 & E54 S12 & TRACE & 195 & \checkmark \\
53 & 20031104  & 19:45 - 23:45 & 10486 & X28 & W89 S17 & TRACE & 195 & \checkmark \\
54 & 20061212  & 20:00 - 23:00 & 10930 &  & W21 S05 & XRT & Thin-Be & \checkmark \\
55 & 20061213  & 02:30 - 05:00 & 10930 & X3.4 & W35 S06 & XRT & Thin-Be & \checkmark \\
56 & 20070313  & 01:00 - 11:00 & 10946 & & W86 N07 & XRT & Ti-Poly & \\
57 & 20070509  & 03:00 - 06:00 & 10953 &  & W91 S11 & XRT & Ti-Poly & \checkmark \\
58 & 20070520  & 19:45 - 20:30 & 10956 &  & W21 N02 & XRT & Ti-Poly & \checkmark \\
59 & 20071217  & 06:00 - 10:00 & 10978 & C2.2 & W79 S10 & XRT & C-Poly &  \\
60 & 20071218  & 06:00 - 10:00 & 10978 &  & W91 S09 & XRT & C-Poly &  \\
61 & 20080409  & 08:00 - 18:00 & 10989 &  & W90 S18 & XRT & Al-Poly & \checkmark \\
62 & 20100613  & 02:00 - 05:30 & 11081 &  & W72 N24 & XRT & Ti-Poly & \checkmark \\
\hline
\bottomrule \\
\end{tabular*}
\end{sidewaystable}
\end{center}

\begin{center}
\begin{table}
\caption{List of analyzed flares.}
\label{sads_list_analyzed}
\footnotesize
\begin{tabular*}{\textwidth}{@{\extracolsep{\fill}} l  c  c  c  c  c  c  c}
\\
\toprule
{\textbf{\#}} & \textbf{YYYYMMDD} & {\textbf{Instrument}} & {\textbf{\# of Flows}} & {\textbf{SADs}} & {\textbf{SADLs}}  & {\textbf{Limb}} & {\textbf{Disk}} \\
\midrule
\midrule
1 & 19980423  & SXT & 10 & \checkmark &  & E &  \\
2 & 19980427  & SXT & 7 & \checkmark &  & E &  \\
3 & 19980506  & SXT & 8 &  & \checkmark & W &  \\
4 & 19980509  & SXT & 3 & \checkmark &  & W &  \\
5 & 19980816  & SXT & 8 & \checkmark &  & E &  \\
6 & 19980930  & SXT & 5 & \checkmark &  & W &  \\
7 & 19981123  & SXT & 5 & \checkmark &  & E &  \\
8 & 19981223  & SXT & 7 &  & \checkmark & E &  \\
9 & 19990120  & SXT & 25 & \checkmark &  & E &  \\
10 & 19990503  & SXT & 14 & \checkmark &  & E &  \\
11 & 19990725  & SXT & 6 & \checkmark &  & W &  \\
12 & 19991128  & SXT & 6 & \checkmark &  & W &  \\
13 & 20000101  & LASCO & 11 & \checkmark &  & E &  \\
14 & 20000222  & SXT & 6 & \checkmark &  & E &  \\
15 & 20000712  & SXT & 7 &  & \checkmark & E &  \\
16 & 20000712  & SXT & 10 & \checkmark &  & W &  \\
17 & 20001016  & SXT & 6 &  & \checkmark & W &  \\
18 & 20001108  & SXT & 2 & \checkmark &  & W &  \\
19 & 20010402  & SXT & 11 &  & \checkmark & W &  \\
20 & 20010403  & SXT & 5 & \checkmark &  & E &  \\
21 & 20010626  & SXT & 9 & \checkmark &  & E &  \\
22 & 20010927  & SXT & 4 &  & \checkmark & W &  \\
23 & 20011009  & SXT & 2 & \checkmark &  &  & \checkmark \\
24 & 20011030  & SXT & 3 & \checkmark &  & E &  \\
25 & 20011101  & SXT & 4 & \checkmark &  & E &  \\
26 & 20011109  & SXT & 2 & \checkmark &  & W &  \\
27 & 20020421  & TRACE & 48 & \checkmark & \checkmark & W &  \\
28 & 20020723  & TRACE & 53 &  & \checkmark & E &  \\
29 & 20031104  & TRACE & 60 &  & \checkmark & W &  \\
30 & 20061212  & XRT & 6 & \checkmark &  &  & \checkmark \\
31 & 20061213  & XRT & 6 & \checkmark &  &  & \checkmark \\
32 & 20070509  & XRT & 2 & \checkmark &  & W &  \\
33 & 20070520  & XRT & 1 & \checkmark &  &  & \checkmark \\
34 & 20080409  & XRT & 16 &  & \checkmark & W &  \\
35 & 20100613  & XRT & 9 & \checkmark &  & W &  \\
\hline
\bottomrule \\
\end{tabular*}
\end{table}
\end{center}

\end{document}